\documentclass[namedreferences]{solarphysics}
\usepackage[hyperref,optionalrh,solaromanenum]{spr-sola-addons} % For Solar Physics 
\usepackage{graphicx}                    % For eps figures, newer & more powerful
\usepackage{epstopdf}
\usepackage{color}                       % For color text: \color command
\usepackage{breakurl}                         % For breaking URLs easily trough lines
\usepackage{tikz}
\usepackage{verbatim}

\newcommand{\rsun}{{\textrm{R}_\odot}} 
\begin{document}

\begin{article}

\begin{opening}

\title{Investigation of the Middle Corona with SWAP and a Data-Driven Non-Potential Coronal Magnetic Field Model}

%%%%%%%%%%%%%%%%%%%%%%%%%%%%%%%%%%%%%%%%%%%%%%%%%%%
%% Authors Names
%
\author[addressref={aff1,aff2},corref,email={kmeyer001@dundee.ac.uk}]{\inits{K.A.}\fnm{Karen A.}~\lnm{Meyer}\orcid{0000-0001-6046-2811}} 
\author[addressref={aff3},corref,email={}]{\inits{D.H.}\fnm{Duncan~H.}~\lnm{Mackay}\orcid{0000-0001-6065-8531}}
\author[addressref={aff4,aff5},corref,email={}]{\inits{D.-C.}\fnm{Dana-Camelia}~\lnm{Talpeanu}\orcid{0000-0002-9311-9021}}
\author[addressref={aff6},corref,email={}]{\inits{L.A.}\fnm{Lisa~A.}~\lnm{Upton}\orcid{0000-0003-0621-4803}}
\author[addressref={aff5},corref,email={}]{\inits{M.J.}\fnm{Matthew~J.}~\lnm{West}\orcid{0000-0002-0631-2393}}
%%%%%%%%%%%%%%%%%%%%%%%%%%%%%%%%%%%%%%%%%%%%%%%%%%%
%% Runningheads
%
\runningauthor{K.A. Meyer et al.}
\runningtitle{Investigation of the Middle Corona}

%%%%%%%%%%%%%%%%%%%%%%%%%%%%%%%%%%%%%%%%%%%%%%%%%%%
%% Affilations 
%% id shold be the same with \author addressref value.
\address[id=aff1]{Mathematics, School of Science \& Engineering, University of Dundee, Nethergate, Dundee DD1 4HN, UK}
\address[id=aff2]{Division of Computing and Mathematics, Abertay University, Kydd Building, Bell Street, Dundee, DD1 1HG, Scotland, UK}
\address[id=aff3]{School of Mathematics and Statistics, University of St Andrews, North Haugh, St Andrews, KY16 9SS, Scotland, UK}
\address[id=aff4]{Centre for mathematical Plasma Astrophysics (CmPA), Department of Mathematics, KU Leuven, Celestijnenlaan 200B, 3001, Leuven, Belgium}
\address[id=aff5]{SIDC - Royal Observatory of Belgium (ROB), Ringlaan 3, 1180, Brussels, Belgium}
\address[id=aff6]{Space Systems Research Corporation, Alexandria, VA, 22314, USA}

%%%%%%%%%%%%%%%%%%%%%%%%%%%%%%%%%%%%%%%%%%%%%%%%%%%
%%% Abstract 
\begin{abstract}
The large field-of-view of the \emph{Sun Watcher using Active Pixel System detector and Image Processing} (SWAP) instrument on board the \emph{PRoject for Onboard Autonomy 2} (PROBA2) spacecraft provides a unique opportunity to study extended coronal structures observed in EUV in conjunction with global coronal magnetic field simulations. A global non-potential magnetic field model is used to simulate the evolution of the global corona from 1 September 2014 to 31 March 2015, driven by newly emerging bipolar active regions determined from \emph{Helioseismic and Magnetic Imager} (HMI) magnetograms. We compare the large-scale structure of the simulated magnetic field with structures seen off-limb in SWAP EUV observations. In particular, we investigate how successful the model is in reproducing regions of closed and open structures; the scale of structures; and compare the evolution of a coronal fan observed over several rotations. The model is found to accurately reproduce observed large-scale off-limb structures. When discrepancies do arise they mainly occur off the east solar limb due to active regions emerging on the far side of the Sun, which cannot be incorporated into the model until they are observed on the Earth-facing side. When such ``late'' active region emergences are incorporated into the model, we find that the simulated corona self-corrects within a few days, so that simulated structures off the west limb more closely match what is observed. Where the model is less successful, we consider how this may be addressed, through model developments or additional observational products.
\end{abstract}

%%%%%%%%%%%%%%%%%%%%%%%%%%%%%%%%%%%%%%%%%%%%%%%%%%%
%% Keywords
%

\end{opening}
%-------------------------------------------------

%%%%%%%%%%%%%%%%%%%%%%%%%%%%%%%%%%%%%%%%%%%%%%%%%%%
%% Sections
%
\section{Introduction}\label{s:intro} 

The \emph{Sun Watcher using Active Pixel System detector and Image Processing}, (SWAP: \citealp{seaton2013instr,halain2013}) instrument on board the \emph{PRoject for Onboard Autonomy 2} (PROBA2) spacecraft provides a wide field-of-view of the Sun in EUV images from an Earth viewing perspective, allowing for the study of extended, large-scale coronal structures \citep{seaton2013}. With its viewing extent of 1.7$\,\rsun$ along the image axes and 2.5$\,\rsun$ along the diagonals \citep{seaton2013}, SWAP bridges the gap between high-resolution EUV images such as from the \emph{Atmospheric Imaging Assembly} (AIA: \citealp{lemen2012}) on board the \emph{Solar Dynamics Observatory} (SDO), and extended white-light observations from coronographs such as the \emph{Large Angle Spectrometric Coronagraph} (LASCO: \citealp{brueckner1995}) instrument on board the \emph{Solar and Heliospheric Observatory} (SOHO), providing a view of the \emph{middle corona} (e.g. \cite{byrne2014}).

SWAP observations have been used to study large-scale coronal structures such as streamers/pseudostreamers \citep{rachmeler2014,goryaev2014,guennou2016}, prominence cavity regions \citep{bazin2013}, post-flare giant arches \citep{west2015}, and coronal mass ejections \citep{ohara2019}. Observations of such large-scale structures, in particular, persistent structures such as streamers/pseudostreamers and coronal fans \citep{koutchmy2002,morgan2007} can help validate and inform the development of global coronal magnetic field models. Potential field source surface (PFSS) extrapolations are often used to aid in the interpretation of observed structures; for example \cite{goryaev2014} used a PFSS extrapolation to consider the coronal magnetic field structure off the solar limb in comparison with coronal streamer observations.

A PFSS extrapolation provides a force-free and current-free estimate of the coronal magnetic field, typically constructed from a single synoptic magnetogram of the global photospheric magnetic field. In the present study, we simulate the global solar corona using a non-linear force-free field model, which produces a continual evolution of the coronal field over many months, allowing for a ``memory'' of magnetic connectivity and the build up of electric currents and free magnetic energy \citep{mackay2006,yeates2008,mackay2016}.

The non-linear force-free field model simulates the global coronal magnetic field out to 2.5$\,\rsun$, so SWAP observations are ideal for comparison between the structure of the simulated global corona and persistent structures seen in EUV. Under the assumption of a low-$\beta$ coronal plasma (magnetic pressure is dominant over plasma pressure), the plasma is assumed to be largely structured by the magnetic field; therefore we compare coronal structures observed in EUV to magnetic structures within the model. In the present article, we focus in particular on large-scale structures observed off-limb with SWAP, such as regions of open/closed structure, streamers/pseudostreamers, and fans. We examine how well these structures are reproduced by the simulation coronal magnetic field. Where the model performs less successfully, we discuss possible reasons for this, and consider how inaccuracies may be addressed, through model developments or additional observational products.

The article is structured as follows: Sect.~\ref{s:model} describes the global coronal magnetic field model; Sect.~\ref{s:data} describes the observational data used for comparison with the simulation; results of the comparison are presented in Sect.~\ref{s:results}; and Discussion and Conclusions in Sect.~\ref{s:conc}.

\section{Model}\label{s:model} 
A global non-potential magnetic field model is used to simulate the evolution of the Sun's corona from
1 September 2014 until 31 March 2015. This period was chosen for analysis as it is during solar maximum, so there are more and brighter extended coronal structures for comparison with the simulation, allowing us to make the most of SWAP’s wide field of view and the simulation’s height range. It is also the most accurately modelled period by this technique to date, in terms of the accuracy of the bipole input data used to drive the simulation.

{{There are two reasons why the bipole data used in this simulation is more accurate than previous simulations: i) synoptic data from the 
Advective Flux Transport (AFT) model \citep{upton2014b,upton2014a,hathaway2016,upton2018} were used to identify newly emerged active region bipoles. This synoptic data includes new emerging bipoles on the Earth-facing side of the Sun on their date of emergence, both pre- and post-central meridian passage. This is in contrast to previously used synoptic data that only includes bipoles as they cross central meridian (i.e. to be included they must emerge pre-central meridian). This means that bipoles that emerge post-central meridian passage can be identified and included in the present 3D simulations at the correct time.  
The bipoles within the AFT model are determined from \emph{Helioseismic and Magnetic Imager} (HMI) magnetogram observations from the \emph{Solar Dynamics Observatory} (SDO/HMI: \citealp{schou2012}).
ii) A careful day-by-day analysis of the AFT synoptic data was carried out to identify all new bipoles, which, while time-consuming, is more robust than fully automated techniques that we have at the present time.}}

A combination of magnetic flux transport \citep{sheeley2005} and magnetofrictional relaxation simulations  
\citep{vanballe2000,mackay2006} are used.  The magnetic flux transport model simulates the 
evolution of the radial component of the magnetic field $B_r$ at the solar surface under the effects of differential rotation \citep{snodgrass1983}, meridional flow \citep{duvall1979}, surface diffusion representing the effects of convection at the surface, \citep{leighton1964}, and flux emergence. 
The coronal evolution is coupled to the magnetic flux transport model, using the
magnetofrictional model \citep{yang1986}, which simulates a quasi-static evolution of the coronal
magnetic field. The effect of this coupling is a continuous evolution of the coronal magnetic field through a series of equilibria, as the magnetofrictional model acts to relax the magnetic field towards a non-linear force-free state, in response to the combined effects of surface transport and flux emergence. The models are coupled at the level of the photosphere, where the evolution of the radial component of the magnetic field is provided as a lower boundary condition to the coronal model. The model does not include a chromosphere or transition region, so we make the assumption of a low-$\beta$ plasma from the first grid point within the corona, which is approximately at 11,500\,km for the resolution used in this simulation.

The Sun's large-scale magnetic field, ${\mathbfit{B}} = (B_r,B_\theta,B_\phi) = \nabla 
\times {\mathbfit{A}}$ is evolved forward in time through the magnetic induction equation, where $(r,\ \theta,\ \phi)$  are radius, colatitude, and longitude. The simulation resolution is $56\times180\times360\; (r,\ \theta,\ \phi)$. It is periodic in $\phi$, with closed boundaries at $\theta=\pm89^\circ$ and $r=\rsun$ and an open boundary at $r=2.5\,\rsun$. We impose the condition that the magnetic field is radial at $r=2.5\,\rsun$. {{The source surface is placed at $2.5\,\rsun$ for two reasons. The first is that within the inner regions of the corona ($r<2.5\,\rsun$) the magnetic pressure is generally much greater than the gas pressure (plasma $\beta<<1$), so below this radial height magnetic forces are dominant and the force-free approximation is valid \citep{mackay2006}. Secondly, comparison of white light eclipse images with magnetic field models has shown that setting the magnetic field to be purely radial at $r=2.5\,\rsun$ gives a good fit between the shape of the magnetic field and the observed corona \citep{altschuler1969}.}} The radial component of the magnetic field at the photosphere $B_r$ is evolved via the induction equation at $r=\rsun$, in terms of the horizontal components of the vector potential ($A_\theta$ and $A_\phi$):
\begin{eqnarray}
\frac{ \partial A_\theta}{\partial t} &=& u_\phi B_r - \frac{D}{r \sin \theta}
\frac{\partial  B_r}{\partial \phi} + S_\theta(\theta, \phi,t)   \label{eq:eqn1}\\
\frac{ \partial A_\phi}{\partial t} &=& -u_\theta B_r + \frac{D}{r} \frac{\partial
B_r}{\partial \theta} + S_\phi(\theta, \phi,t), \label{eq:eqn2}
\end{eqnarray}
where $D$ is the
photospheric diffusion constant \citep[$D = 450$\,km$^2$\,s$^{-1}$; see][]{devore1985},
$u_\phi$ is the azimuthal velocity, and $u_\theta$ the meridional flow velocity. 
The azimuthal velocity is given by
\begin{eqnarray}
u_\phi = \Omega(\theta) r \sin \theta,% \nonumber
\end{eqnarray}
where $\Omega(\theta)$ is the angular velocity of differential rotation
relative to the Carrington frame \citep{snodgrass1983},
\begin{eqnarray}\label{eqn:DR}
\Omega(\theta) = 0.18 - 2.30 \cos^2 \theta - 1.62 \cos^4 \theta \hspace*{0.1cm} {\rm deg\,day}^{-1}.
\end{eqnarray}
The poleward meridional flow is given by
\begin{eqnarray}\label{eqn:merid}
u_\theta = -C \sin( 2 \lambda) \,\exp( \pi - 2 \mid \lambda \mid)
\end{eqnarray}
where $C = 15$ m s$^{-1}$  and $\lambda = \frac{\pi}{2} - \theta$ \citep{schussler2006}. The terms $S_{\theta}$ and 
$S_{\phi}$ are source terms representing the emergence of new magnetic flux. To specify the
source term for the emergence of new magnetic flux, daily synoptic magnetograms produced from the AFT model
%Advective Flux Transport (AFT) model \citep{upton2014b,upton2014a,hathaway2016,upton2018}
are studied and a semi-automated method for identifying the emergence of new bipoles used (see Appendix A of \cite{yeates2018} for details). 
Once new emerging bipoles are identified, their properties such as day of peak flux, longitude, latitude, size, and tilt angle are determined. The emergence of each bipole is then simulated on its day of peak flux.

The coronal magnetic field evolves in response to photospheric boundary motions, via the induction equation,                                                              
\begin{eqnarray}                                                                           
\frac{ \partial {\mathbfit{A}}}{ \partial t} = {\mathbfit{v}} \times {\mathbfit{B}}, %\label{eq:cor}
\end{eqnarray}                                                                                   
where ${\mathbfit{v}}({\mathbfit{r}},t)$ is the plasma velocity. The coronal plasma velocity is assumed to be
\begin{eqnarray}                                                                                 
{\mathbfit{ v}} = \frac{1}{\nu} \frac{ {\mathbfit{j}} \times {\mathbfit{B}} }{ B^2} +                               
v_o \mathrm{e}^{-(2.5\small\rsun- r)/r_w} {\hat\mathbfit{ r}}. %\nonumber                                            
\end{eqnarray} 
where ${\mathbfit{j}} = \nabla \times {\mathbfit{B}}$, and $\nu$ is the coefficient of friction.
 The first term on the right hand side is the magnetofrictional velocity \citep{yang1986} and 
reflects the fact that the Lorentz force is dominant in the corona (low-$\beta$ condition). 
The effect of this ``frictional'' term is that, wherever the coronal magnetic field departs from a force-free state, e.g. as a
result of lower (photospheric) boundary motions,  magnetic forces in the corona act to restore the field towards a 
force-free state (generally, a non-linear force-free field). The second term represents a radial outflow 
velocity, which is imposed to ensure that the field lines remain radial at the source surface ($r=2.5\,\rsun$). This outflow velocity approximates
 the effect of the solar wind in opening coronal field lines. Its peak value is set to be
$v_o = 100$\,km\,s$^{-1}$ and its exponential fall-off length from the outer boundary is
$r_w = 0.1\,\rsun$. This term is negligible in the low closed-field corona. A peak value of 100\,km\,s$^{-1}$ was chosen for the radial outflow velocity as previous studies have shown that it is the optimal value for i) removing flux rope eruptions from the top of the computational box by advecting them through the top boundary, while ii) producing an insignificant effect on the closed magnetic field lower down in the corona \citep{mackay2006}. Full details of the computational grid used
can be found in the articles of \cite{mackay2006}, \cite{yeates2008}, \cite{yeates2012}, and \cite{mackay2016}.

The initial photospheric magnetic field distribution over the whole Sun is taken from the AFT model's daily synoptic map from 1 September 2014. This map produces the best estimate for the magnetic field on the start day for both the visible side and the far side. The initial condition for the global coronal magnetic field is a PFSS extrapolation out to 2.5\,$\rsun$ on the same date. The coronal field is then continually evolved for 200 days, allowing the self consistent build-up of electric currents and free magnetic energy, in response to photospheric driving (both magnetic footpoint motions and flux emergence). A fully non-potential corona was reached after six to eight weeks \citep{yeates2018}, so the first date that we consider for comparison with observations is 11 October 2014. This simulation has previously been compared with other global non-potential coronal magnetic models for the 20 March 2015 solar eclipse \citep{yeates2018}. In contrast, in the present article, we focus on a comparison of the simulated large-scale corona, with extended off-limb structures observed in
EUV with SWAP.

\section{Data}\label{s:data} 

The SWAP instrument on board PROBA2 is a large field-of-view EUV imager with a passband centered
around 174\,\AA{}, that corresponds to a temperature of roughly one million
degrees. SWAP's nominal observation mode produces an image every one to two
minutes. A separate \emph{stacked} dataset, based on the nominal data
has been developed to help reduce signal-to-noise in the far field of the
images. The images are generated by adding (stacking) and median
filtering all images over a 100 minute period (nominally 40\,--\,50 images).
The effective longer exposure time helps to generate a greater
signal-to-noise in the images, enhancing faint off-limb EUV structures
at greater heights. The FITS images are available at
\verb!proba2.sidc.be/swap/data/carrington_rotations/! or can be
produced using the IDL routine p2sw\_long\_movie.pro in the SWAP SSWIDL
software. The stacked images are further processed using the Multiscale Gaussian Normalization technique of \cite{morgan2014}, to greater enhance the faint off-limb features. These are used for comparison with the simulation
corona. 

SDO/HMI full disc magnetograms are used for comparison with the simulation photospheric magnetic field.
The period from 1 September 2014 to 20 March 2015 was near solar maximum, so the EUV corona was very active, with many extended structures observed off the limb by SWAP (see figures and animations in Sect.~\ref{s:results}). Many large active regions emerged during this time, which can also be seen in the HMI magnetograms.

\section{Results}\label{s:results}

  \begin{figure}
   \centerline{\hspace*{0.0\textwidth}
               \includegraphics[width=0.5\textwidth,clip=]{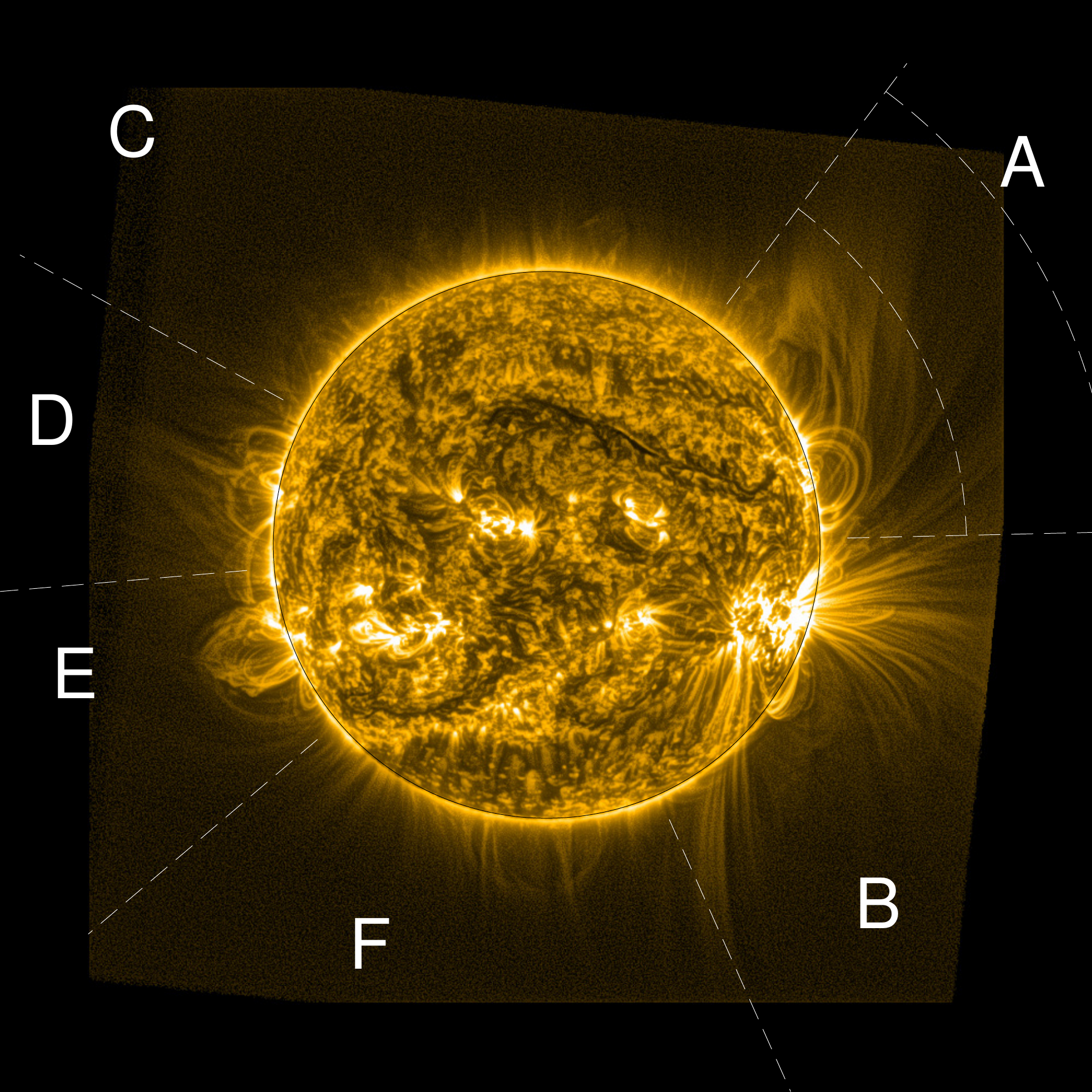}
               \hspace*{-0.0\textwidth}
               \includegraphics[width=0.5\textwidth,clip=]{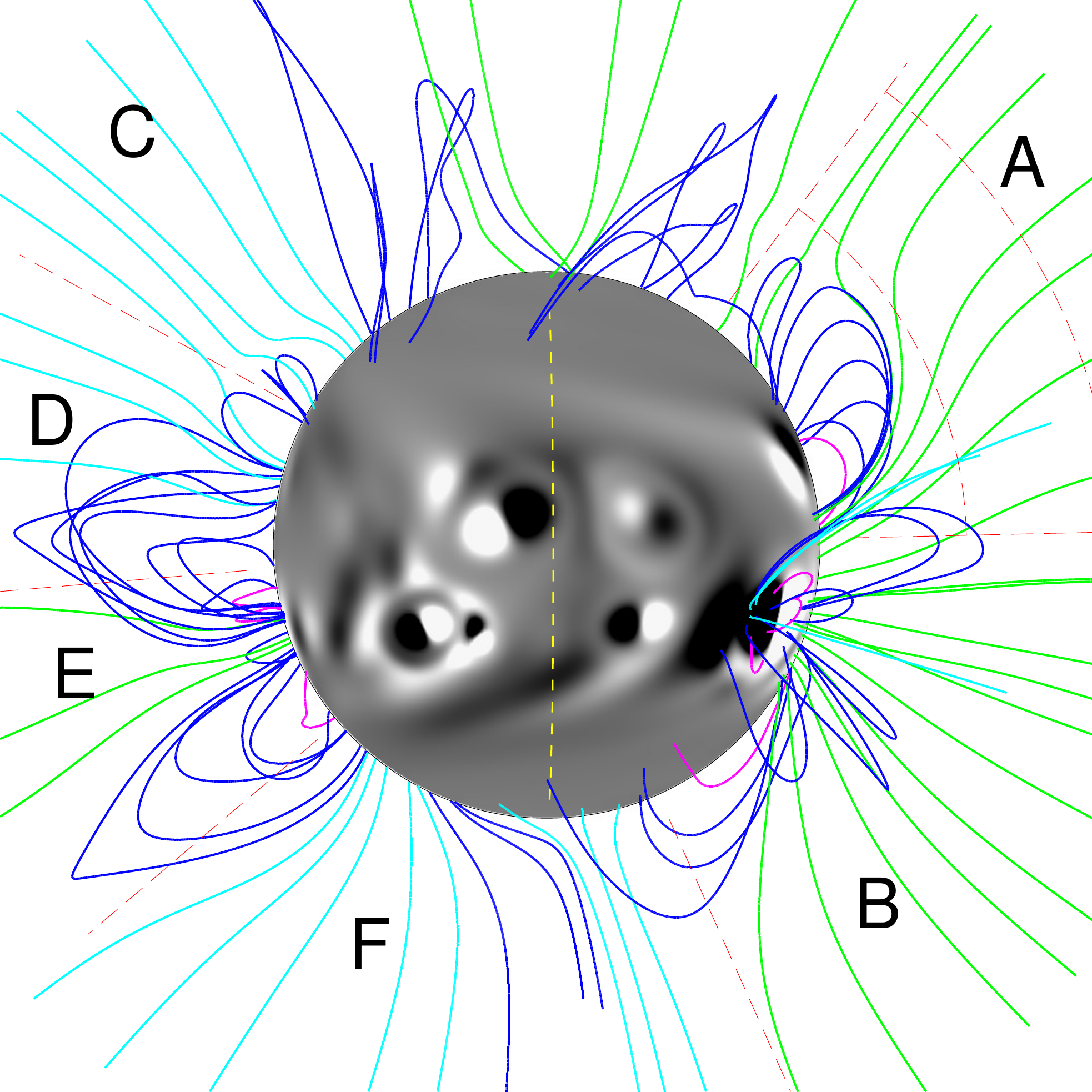}
              }
     \vspace{-0.5\textwidth}   % Shift close to the panel top 
     \centerline{\Large \bf     % Includes the labels (here needs the color 
                                %   package, see beginning of this file)
      \hspace{-0.1 \textwidth}  \color{black}{(a)}
      \hspace{0.49\textwidth}  \color{black}{(b)}
         \hfill}
     \vspace{0.46\textwidth}    % Shift back to the panel bottom 

\caption{27 October 2014: (a) SWAP EUV image, white dashed lines and letters A\,--\,F indicate zones for comparison. Dashed arcs in zone A are plotted at 0.54\,$\rsun$ and 1.07\,$\rsun$ above the photosphere. (b) Simulation showing the photospheric magnetic field $B_r$ saturated at $\pm$30\,G and a selection of coronal magnetic field lines in magenta (closed, low-lying), dark blue (closed), green (open, positive field), and light blue (open, negative field). Red dashed lines indicate the same regions as (a). Central meridian is indicated by the yellow dashed line.}\label{fig:056}
   \end{figure}

  \begin{figure}
   \centerline{\hspace*{0.0\textwidth}
               \includegraphics[width=0.5\textwidth,clip=]{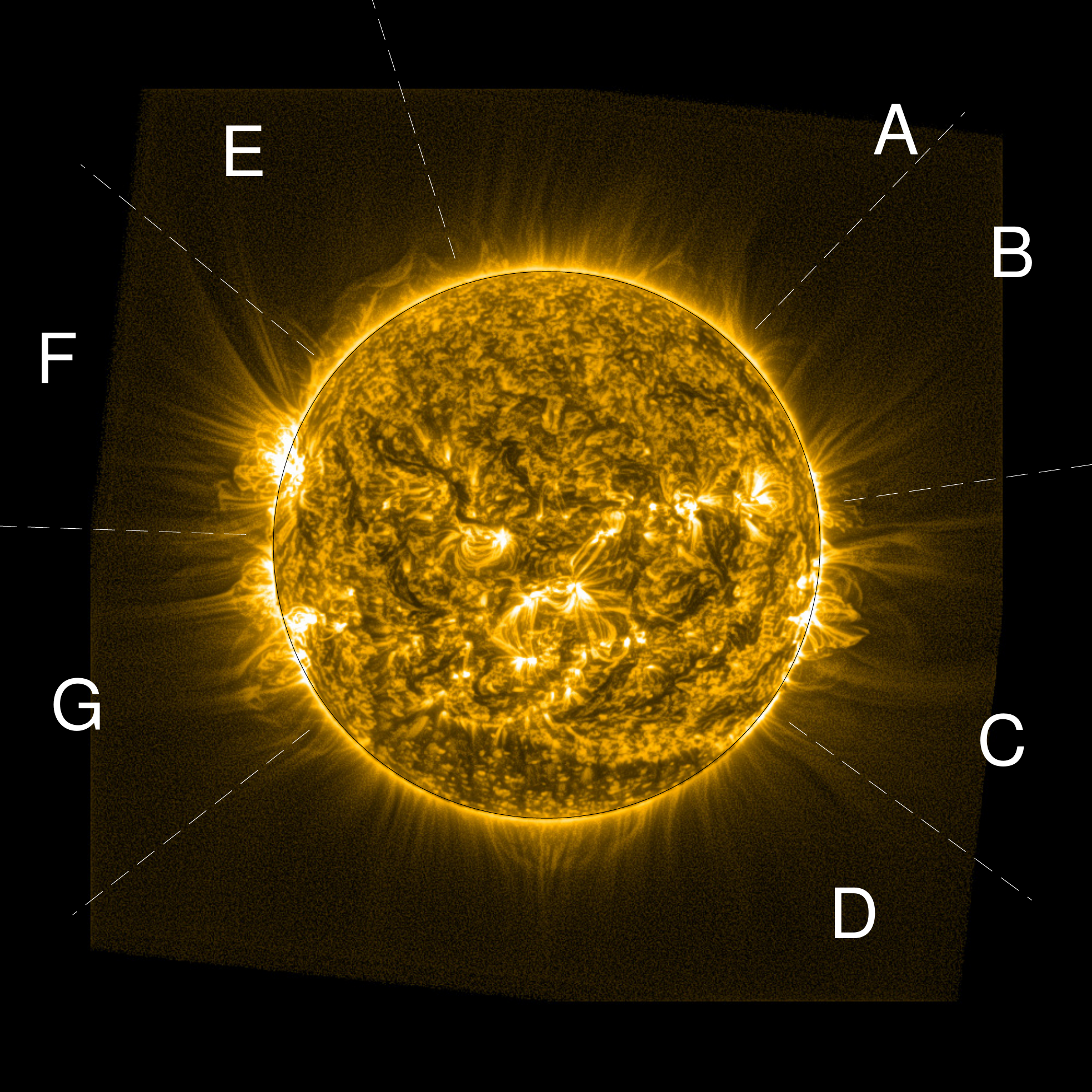}
               \hspace*{-0.0\textwidth}
               \includegraphics[width=0.5\textwidth,clip=]{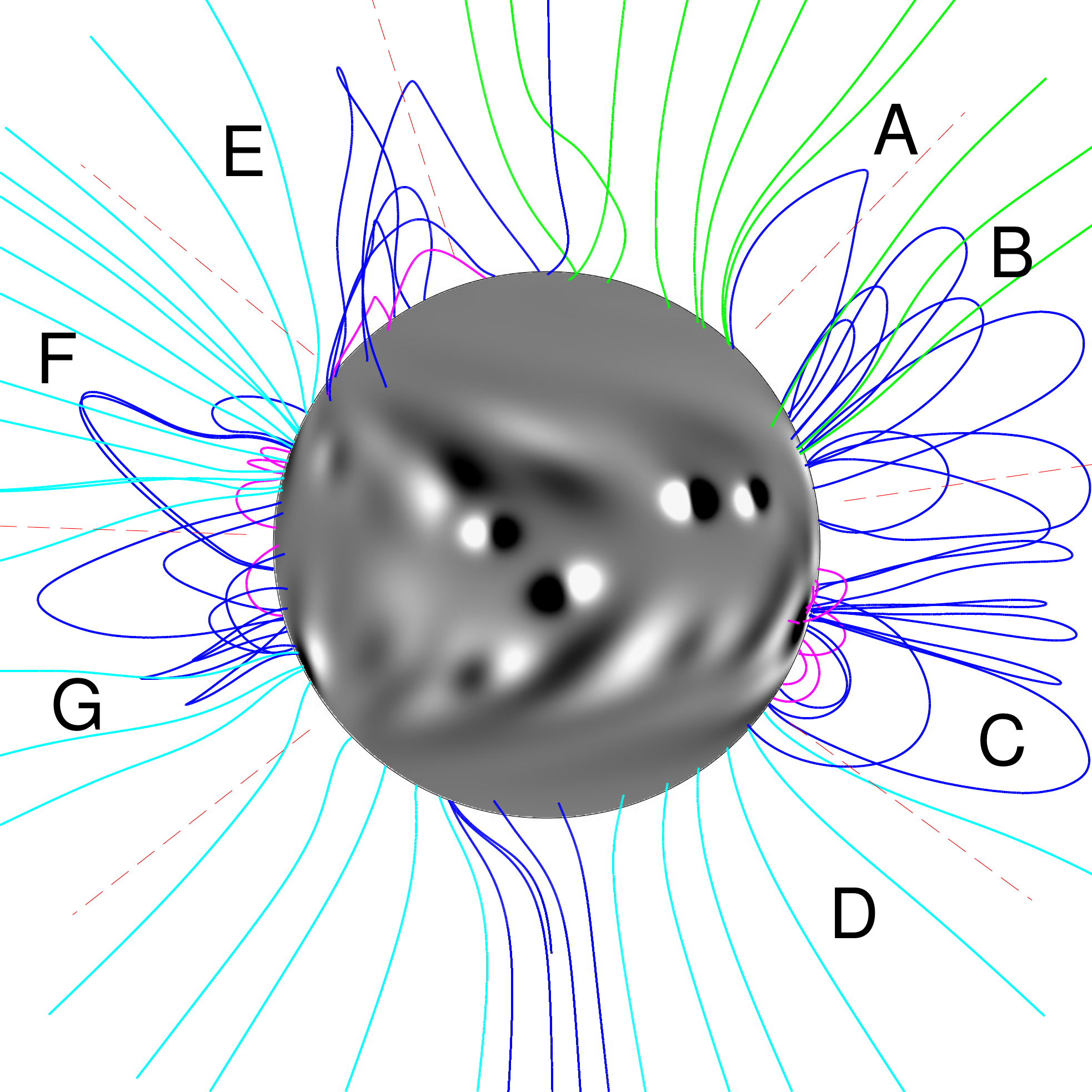}
              }
     \vspace{-0.5\textwidth}   % Shift close to the panel top 
     \centerline{\Large \bf     % Includes the labels (here needs the color 
                                %   package, see beginning of this file)
      \hspace{-0.1 \textwidth}  \color{black}{(a)}
      \hspace{0.49\textwidth}  \color{black}{(b)}
         \hfill}
     \vspace{0.46\textwidth}    % Shift back to the panel bottom 

\caption{4 November 2014: (a) SWAP EUV image, white dashed lines and letters A\,--\,G indicate zones for comparison. (b) Simulation showing photospheric magnetic field $B_r$ saturated at $\pm$30\,G and a selection of coronal magnetic field lines in magenta (closed, low-lying), dark blue (closed), green (open, positive field), and light blue (open, negative field). Red dashed lines indicate the same regions as (a).}\label{fig:064}
   \end{figure}

  \begin{figure}
   \centerline{\hspace*{0.0\textwidth}
		\begin{tikzpicture}
		\begin{scope}
 		   \node[anchor=south west,inner sep=0] (image) at (0,0) {\includegraphics[width=0.5\textwidth]{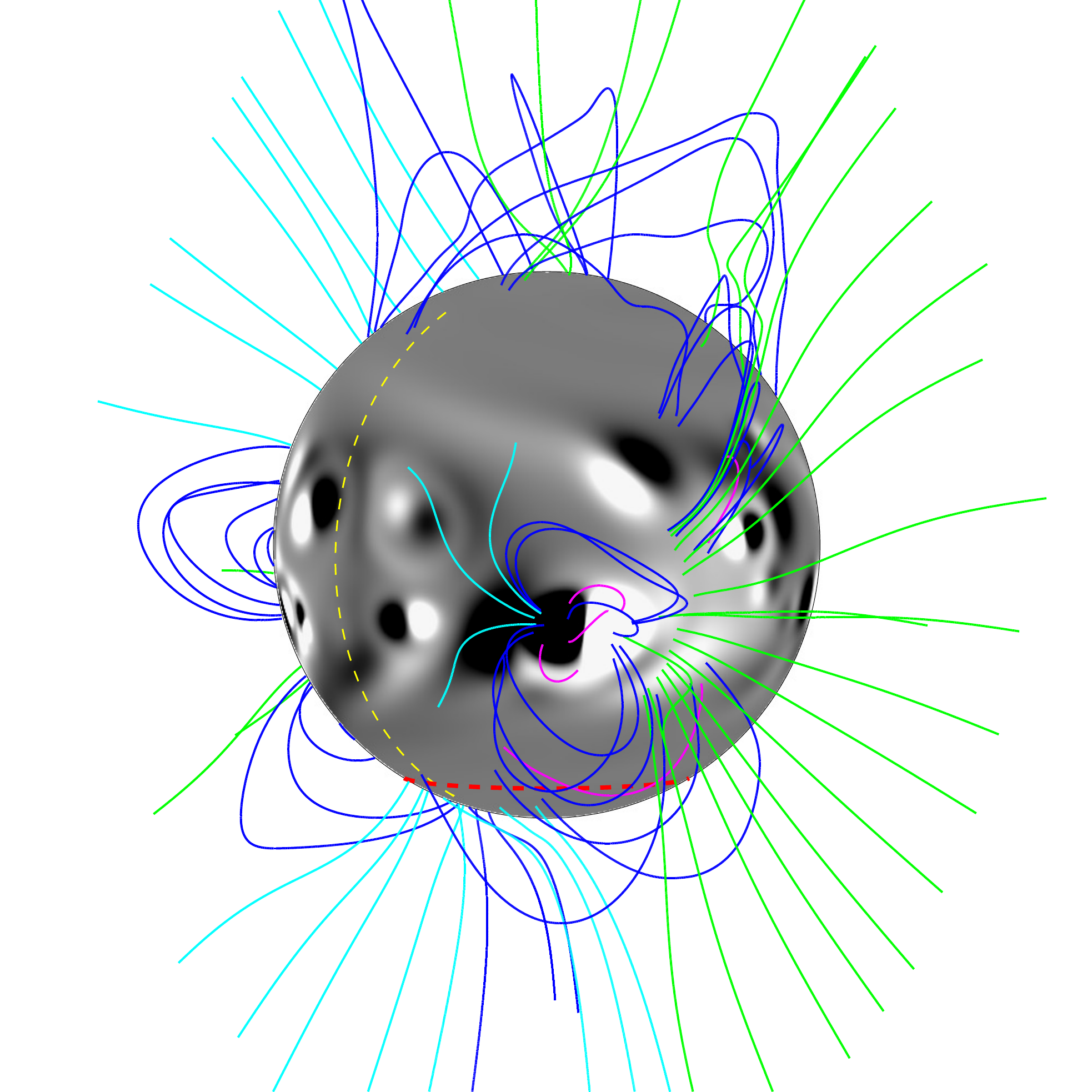}};
  		  \begin{scope}[x={(image.south east)},y={(image.north west)}]
    		    \draw [->, line width=3pt, yellow] (0.45,0.58)--(0.5,0.5);
  		  \end{scope}
		\end{scope}
		\end{tikzpicture}%
               \hspace*{-0.0\textwidth}
               \includegraphics[width=0.5\textwidth,clip=]{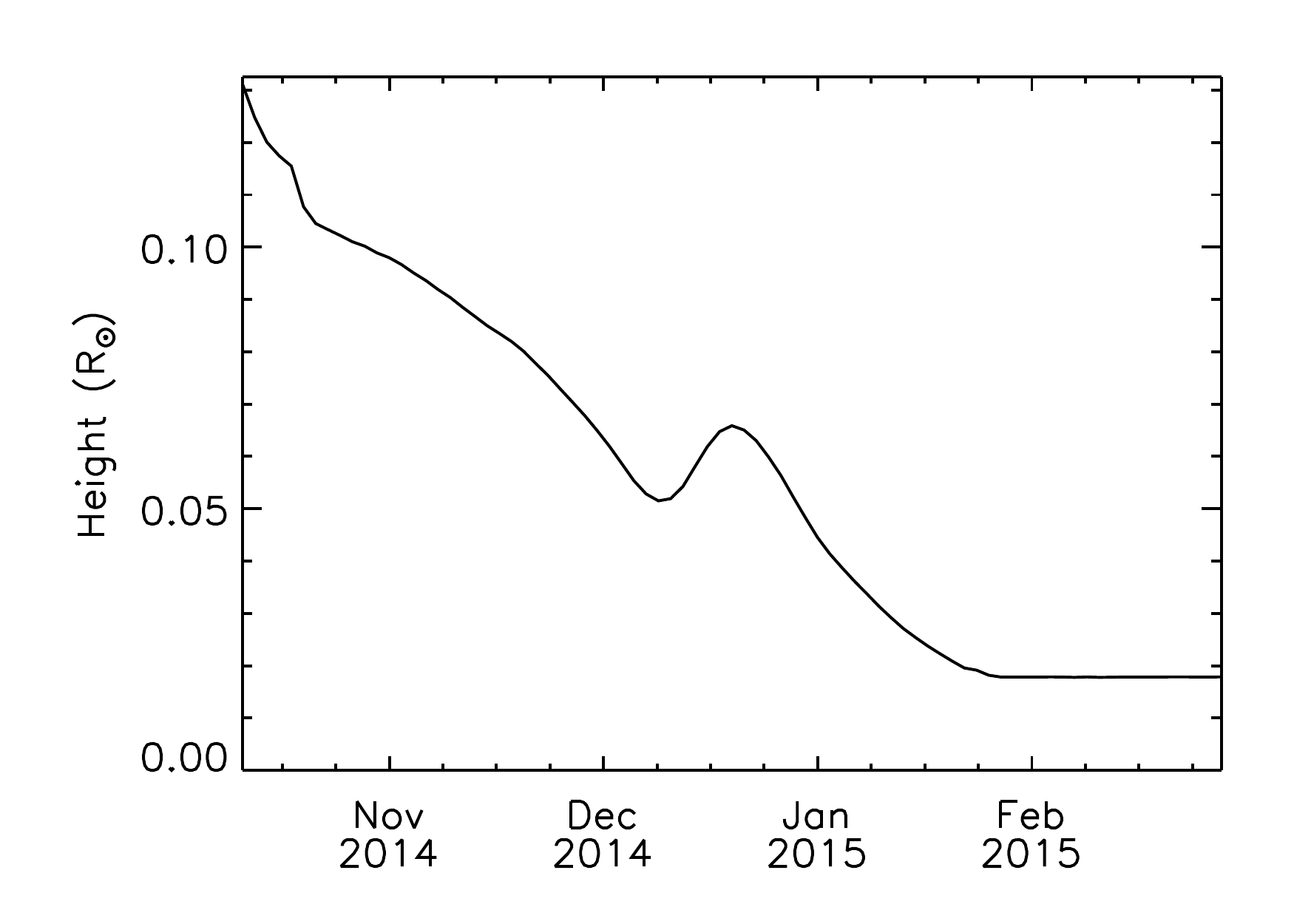}
              }
     \vspace{-0.5\textwidth}   % Shift close to the panel top 
     \centerline{\Large \bf     % Includes the labels (here needs the color 
                                %   package, see beginning of this file)
      \hspace{-0.1 \textwidth}  \color{black}{(a)}
      \hspace{0.49\textwidth}  \color{black}{(b)}
         \hfill}
     \vspace{0.46\textwidth}    % Shift back to the panel bottom 

\caption{27 October 2014: (a) Simulation showing photospheric magnetic field $B_r$ saturated at $\pm$30\,G and a selection of coronal magnetic field lines in magenta (closed, low-lying), dark blue (closed), green (open, positive field), and light blue (open, negative field). The field lines are the same as those plotted in Fig.~\ref{fig:056}b, with the Sun rotated 50 degrees in longitude in the Eastward direction. Central meridian is indicated by the vertical yellow dashed line. The horizontal red dashed line indicates $-76.8$ degrees latitude. The active region seen on the limb in zone B in Figure~\ref{fig:056}b is indicated by the yellow arrow. (b) Maximum height of closed magnetic field structures at the South Pole (latitude$<-76.8$ degrees) within the simulation, versus time.}\label{fig:056a}
   \end{figure}

\subsection{Comparison of EUV Observations with Simulated Magnetic Field}

To begin with, we compare the general structure of the large-scale corona observed by SWAP in EUV (in particular, off-limb structures) with the magnetic field structure produced in the simulation on the same day. Figures~\ref{fig:056} and \ref{fig:064} show the EUV and simulation comparison for 27 October 2014 and 4 November 2014, respectively. The labels A, B, C, etc. between white/red dashed radial lines off the solar limb indicate zones of interest for comparison. The simulation photospheric magnetic field $B_r$ is saturated at $\pm30$\,G and a selection of magnetic field lines are shown to illustrate the coronal structure. Coloured field lines indicate low-lying closed (magenta), large-scale closed (dark blue), open positive (green), and open negative (light blue) magnetic field. The dashed arcs plotted in zone A in Fig.~\ref{fig:056}a and b are situated at 0.54\,$\rsun$ and 1.07\,$\rsun$ above the photosphere, to aid in the comparison of scale. The central meridian, as seen from Earth, is indicated by the yellow dashed line in Fig.~\ref{fig:056}b. Detailed descriptions of these figures are given below. An animation (SWAP\_sim\_compare.mp4) is included in the Electronic Supplementary Materials, showing the simulation and SWAP EUV images side-by-side, once every two days from 11 October 2014 to 20 March 2014.

On 27 October 2014 (Fig.~\ref{fig:056}b), the simulated coronal magnetic field reproduces the general structure of the corona seen in EUV (Fig.~\ref{fig:056}a). In Fig.~\ref{fig:056}a, a fan can be seen off the north-west limb in the EUV image in zone A. The general shape and scale of the fan is reproduced by the simulation magnetic field, as can be seen in Fig.~\ref{fig:056}b. This fan persists for several solar rotations and will be discussed in detail in Sect.~\ref{sec:fans}.

The morphology of the large active region in zone B is also reproduced by the simulation, with open and closed field fanning outward from the solar limb. From the EUV image (Fig.~\ref{fig:056}a), it is not clear whether the open structures seen off the limb in zone B originate from the active region close to the limb, or from behind it. Fig.~\ref{fig:056a}a shows the same simulation date and magnetic field lines as in Fig.~\ref{fig:056}b, but the Sun has been rotated 50 degrees in longitude so that the active region seen on the limb in zone B (Fig.~\ref{fig:056}b) can now be seen on the disc. The central meridian is indicated by the yellow dashed line. From this angle, it can be seen that the open magnetic field in zone B of the simulation originates from behind the active region, from magnetic flux that has resulted from the decay of an older active region. This demonstrates that the model can be used to aid in the interpretation of the observations and in particular the origin of both open and closed field lines.

In Fig.~\ref{fig:056}, active region loops off the east limb (D and E) are present in both the EUV observations and simulation, although the loops extend to much greater heights in the simulation. One reason for this could be that due to the lower density in the upper corona, loops at larger heights may not be visible in the EUV images, even if they do exist. There are closed structures at the North Pole (zone C) in the simulation, but no similar structure is seen in the EUV. The open field at the South Pole (zone F) has a cusp shape to its structure in the EUV, which can also be seen in the simulation. This cusp structure is a large-scale coronal pseudostreamer/cavity system that persisted for approximately one year, from February 2014 until March 2015, and is discussed in detail by \cite{guennou2016}. They observed the open field at the South Pole to be negative during this time, as is also the case in our simulation (light blue open field lines). The pseudostreamer gradually shrinks with time, until it disappears in early March 2015. This can also be seen in the simulation. 

Fig.~\ref{fig:056a}b shows a plot of the maximum height of closed structures at the South Pole versus time, within the simulation. The plot was produced by tracing magnetic field lines originating below $-76.8$ degrees latitude in the simulation, and determining the maximum height that any closed field lines reached. {{The value of $-76.8$ degrees latitude was chosen as it is high enough in latitude to have few large-scale connections to active latitudes, so that we consider only the pole. The specific value of $-76.8$ coincides with a grid point within the model.}} The plot shows that, generally, the simulated maximum height of closed structures at the South Pole decreases with time, levelling off at the end of January 2015. The height of structures in the simulation are less than that of the pseudostreamer observed by \cite{guennou2016}, however. They found that the pseudostreamer extended to around 0.1\,$\rsun$ above the photosphere in December 2014 (see their Figure 1E), whereas the maximum height of closed structures in the simulation South Pole during December is 0.05\,--\,0.07$\,\rsun$. This difference can be attributed in part to the timescale of flux transport via meridional flow (Eq.~\ref{eqn:merid}), which is two years \citep{mackay2012}. Since the initial condition is a potential magnetic field and the simulation only runs for seven months in total, there has not yet been sufficient time for the correct potential/non-potential balance around the polar regions to occur through the self-consistent injection of bipoles and their subsequent transport poleward.

  \begin{figure}

   \centerline{\hspace*{0.0\textwidth}
               \includegraphics[width=0.33\textwidth,clip=]{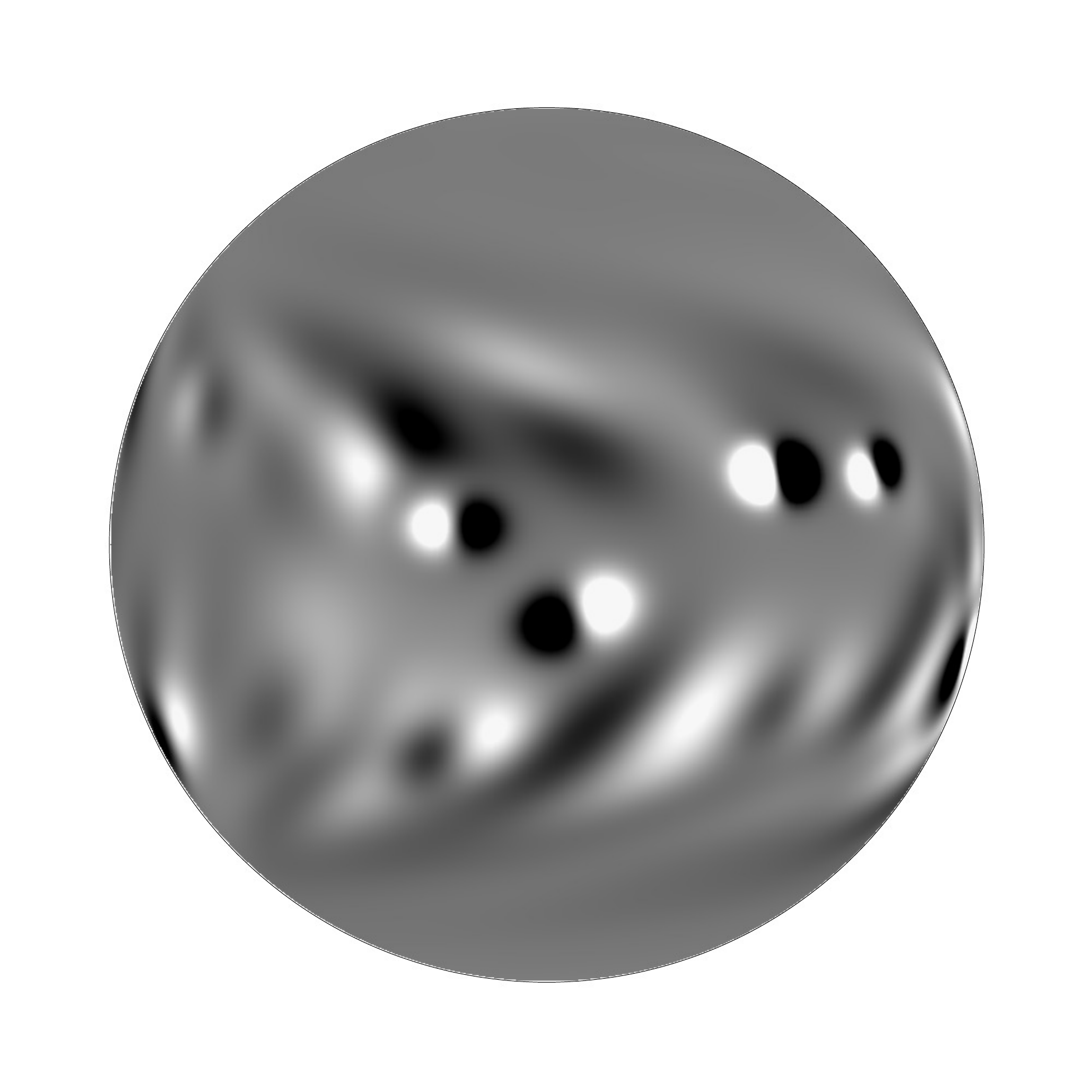}
               \hspace*{-0.0\textwidth}
               \includegraphics[width=0.33\textwidth,clip=]{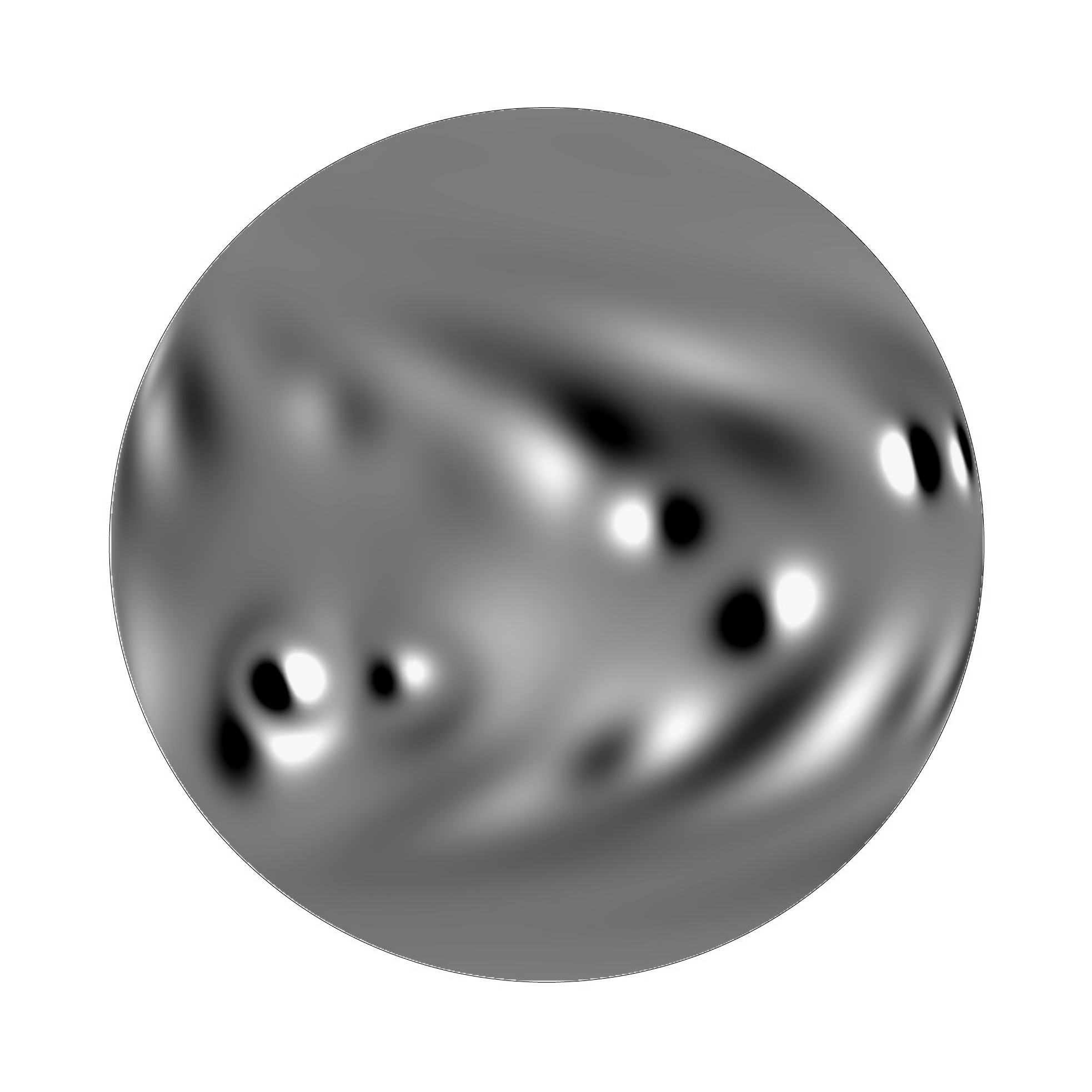}
               \hspace*{-0.0\textwidth}
		\begin{tikzpicture}
		\begin{scope}
 		   \node[anchor=south west,inner sep=0] (image) at (0,0) {\includegraphics[width=0.33\textwidth]{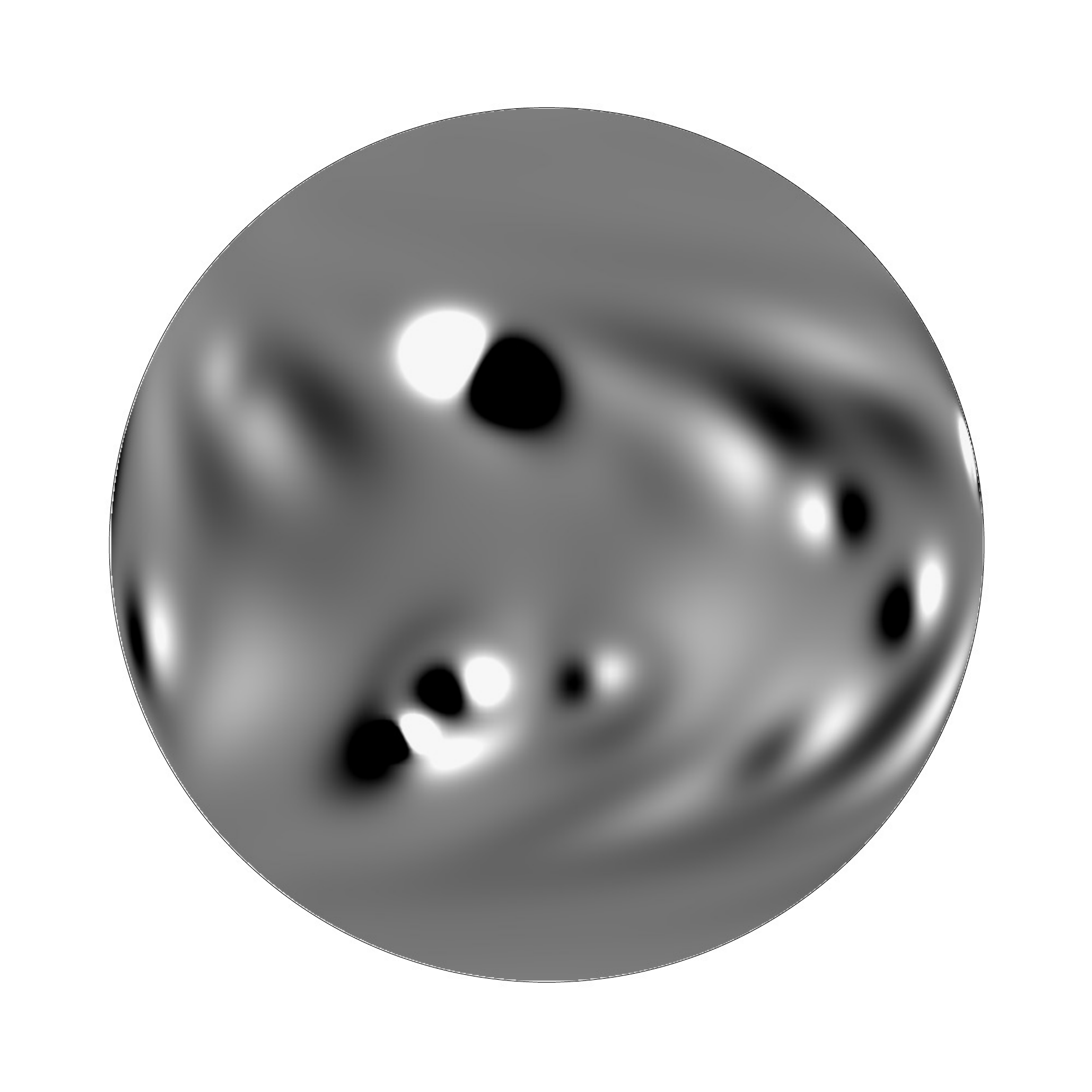}};
   		 \begin{scope}[x={(image.south east)},y={(image.north west)}]
		        \draw [->, line width=3pt, yellow] (0.53,0.85)--(0.46,0.75);
  		  \end{scope}
		\end{scope}
		\end{tikzpicture}%
              }

     \vspace{-0.34\textwidth}   % Shift close to the panel top 
     \centerline{\Large \bf     % Includes the labels (here needs the color 
                                %   package, see beginning of this file)
      \hspace{-0.03 \textwidth}  \color{black}{\small (a)}
      \hspace{0.28 \textwidth}  \color{black}{\small (b)}
      \hspace{0.28 \textwidth}  \color{black}{\small (c)}
         \hfill}
     \vspace{0.3\textwidth}    % Shift back to the panel bottom 

   \centerline{\hspace*{0.0\textwidth}
               \includegraphics[width=0.33\textwidth,clip=]{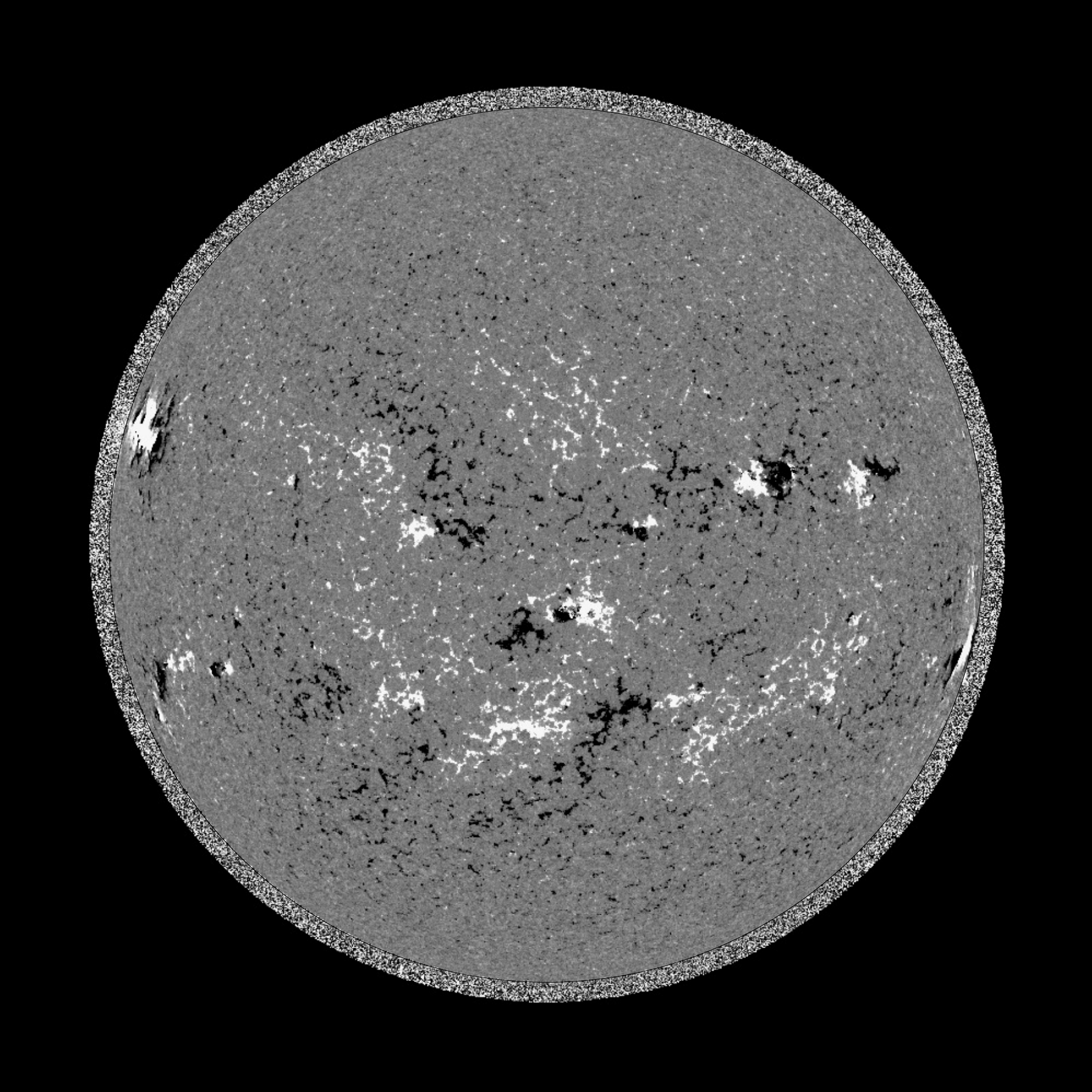}
               \hspace*{-0.0\textwidth}
               \includegraphics[width=0.33\textwidth,clip=]{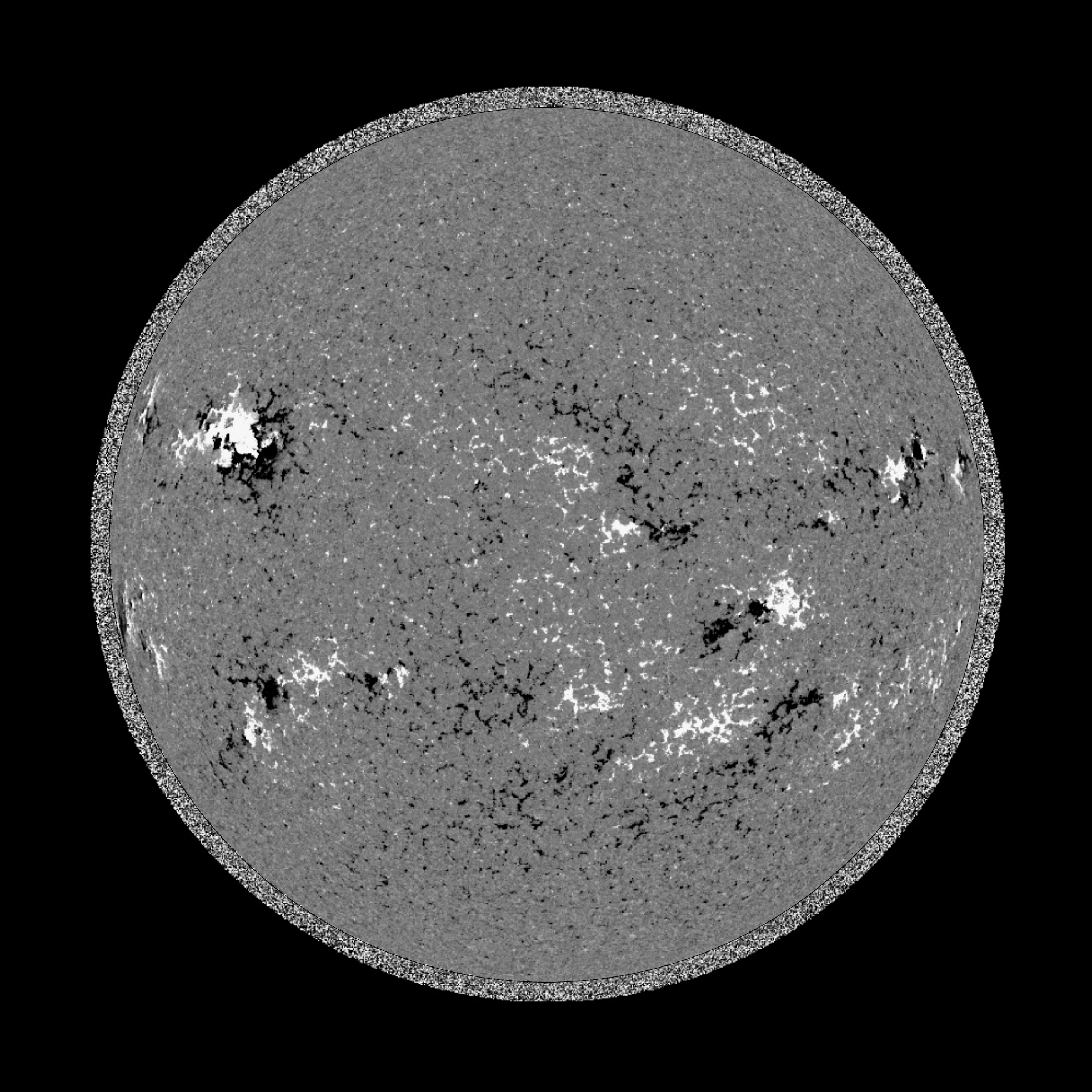}
               \hspace*{-0.0\textwidth}
               \includegraphics[width=0.33\textwidth,clip=]{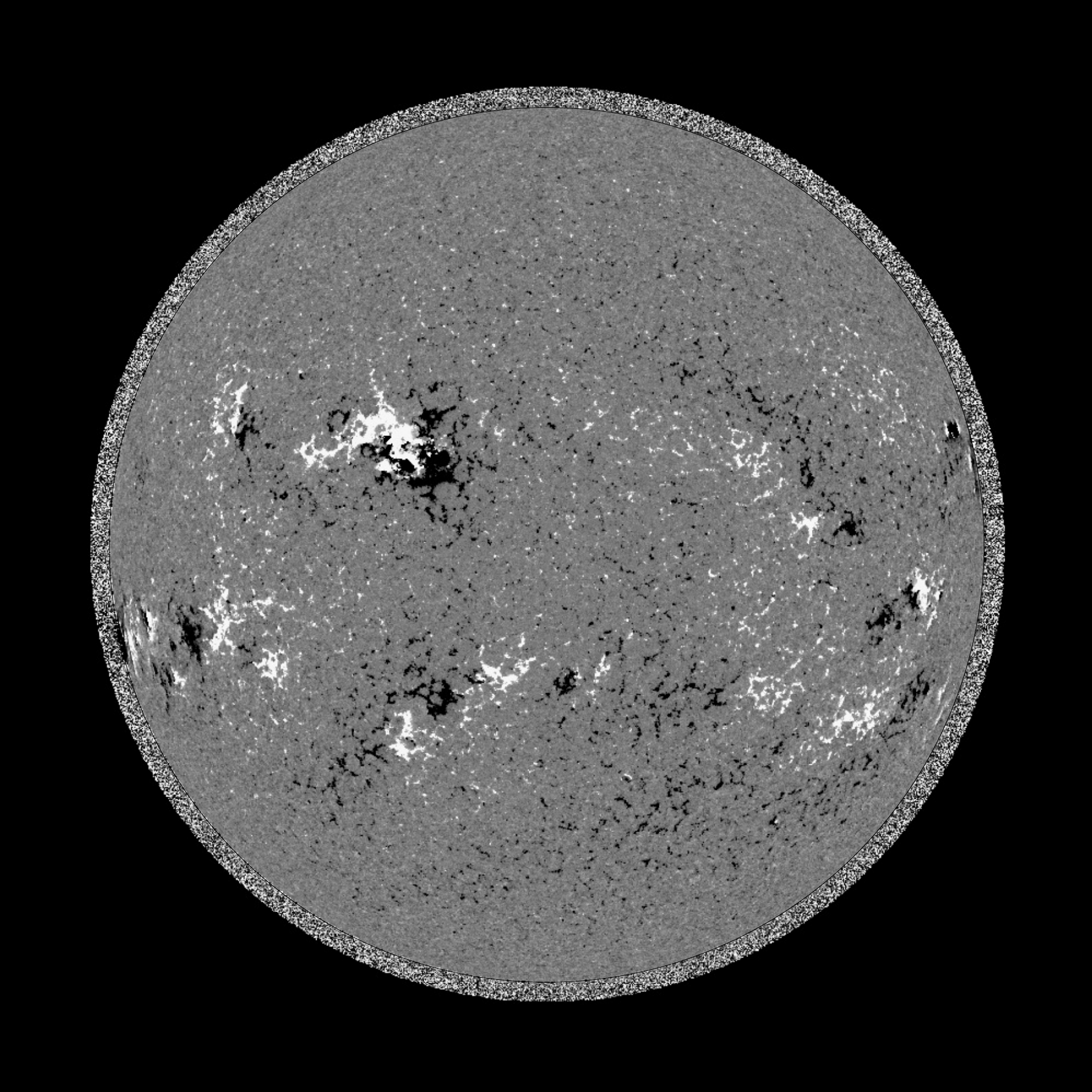}
              }
     \vspace{-0.34\textwidth}   % Shift close to the panel top 
     \centerline{\Large \bf     % Includes the labels (here needs the color 
                                %   package, see beginning of this file)
      \hspace{-0.025 \textwidth}  \color{white}{\small (d)}
      \hspace{0.285 \textwidth}  \color{white}{\small (e)}
      \hspace{0.285 \textwidth}  \color{white}{\small (f)}
         \hfill}
     \vspace{0.31\textwidth}    % Shift back to the panel bottom 

\caption{(a)\,--\,(c) simulation photospheric magnetic field on (a) 4, (b) 6 and (c) 8 November 2014. (d)\,--\,(f) HMI full disc magnetograms on the same dates. All images are saturated at $\pm30$\,G. The yellow arrow indicates an active region that emerged between 6 and 8 November in the simulation.}\label{fig:B}
   \end{figure}

It can be seen in Fig.~\ref{fig:064} that the simulation is able to accurately reproduce some regions of the solar atmosphere, on 4 November 2014. Both low-lying and large-scale active region loops are reproduced in region C; the polar coronal field is predominantly open (A, D); and there is closed field off the north-east limb (E). However, the
simulation does not adequately reproduce the coronal structure off the
east limb. Some of the active regions seen on the east limb in the EUV image (F, G) have not yet  rotated onto the visible disk to be included in the HMI magnetograms and the AFT model, and therefore they have not \emph{yet} been emerged within our simulation. As a result, the simulation's coronal structures here are due to older active region magnetic fields, which have continued to decay while on the far-side of the Sun. 

Figure~\ref{fig:B} shows a comparison between the simulation photospheric magnetic field and full disc photospheric magnetograms observed by HMI on the same day. Images are saturated at $\pm30$\,G. Panels a\,--\,c show the simulation photosphere on 4, 6 and 8 November 2014; panels d\,--\,f shows the same dates for HMI. An animation (HMI\_sim\_compare.mp4) is included in the Electronic Supplementary Materials, showing HMI full disc magnetograms and the simulation photospheric magnetic field side-by-side, once every two days from 11 October 2014 to 20 March 2014.  On  4  November, in the HMI magnetogram (Fig.~\ref{fig:B}d) a large active region can be seen at the east limb, just above the Equator. The bright, coronal plasma of this active region can be seen in the same location in the SWAP EUV image (Fig.~\ref{fig:064}a), but it is not yet present within the simulation (Fig.~\ref{fig:B}a and  Fig.~\ref{fig:064}b). The active region emerged on the far side of the Sun, so it cannot currently be assimilated into the simulation until it is observed on the Earth-facing side. Figure~\ref{fig:B}c shows that the active region has been inserted into the simulation between  6 and 8 November (indicated by yellow arrow). A limitation of the model in its current form is that active regions are chosen to emerge within the simulation on the day of their peak observed flux. Additional magnetogram observations from a spacecraft located at the $L_5$ Lagrange point would also be of benefit for this type of scenario \citep{trichas2015}. This will be discussed further in Sect.~\ref{s:conc}.

\subsection{Coronal Fan Case Study}\label{sec:fans}

  \begin{figure}

   \centerline{\hspace*{0.015\textwidth}
               \includegraphics[width=0.54\textwidth,clip=]{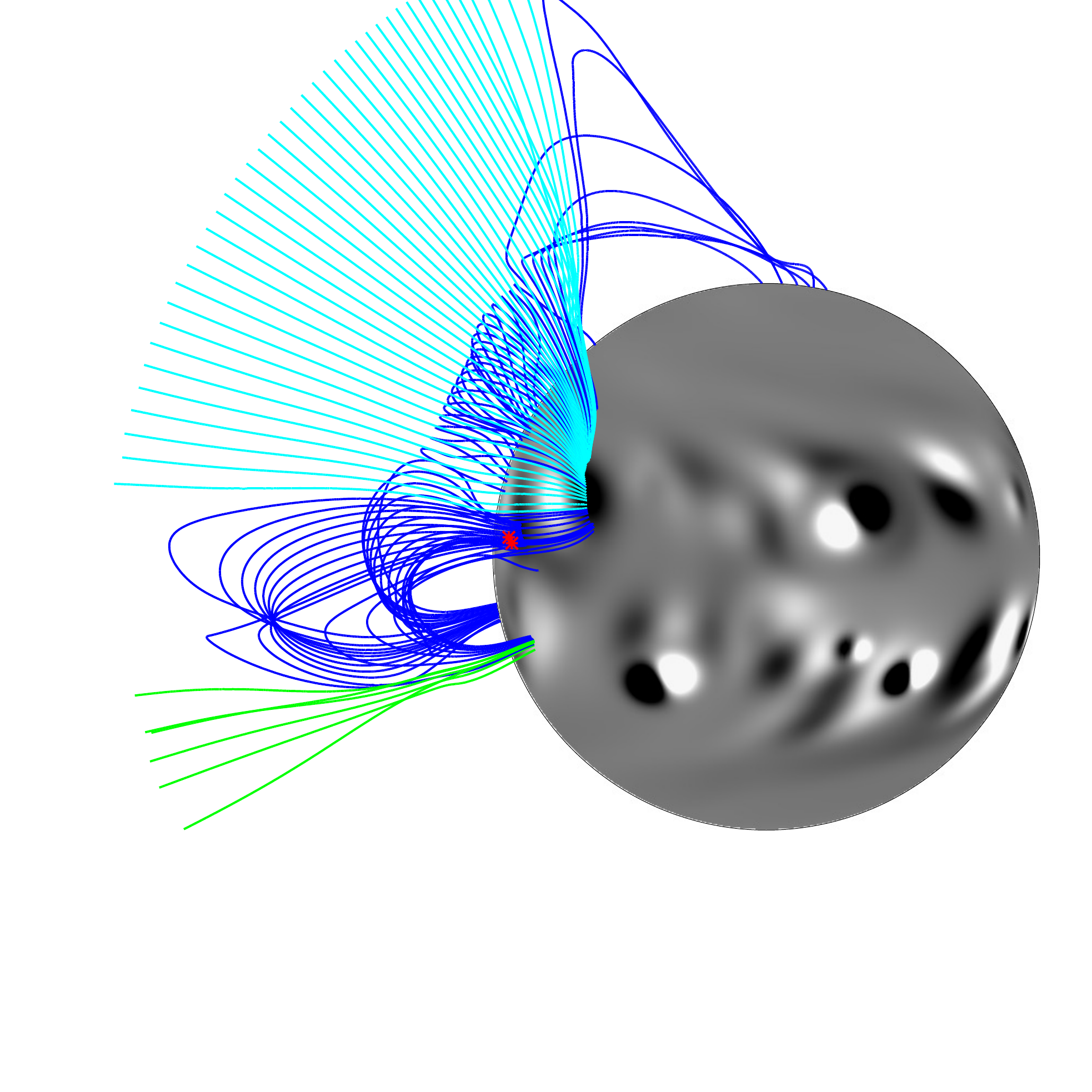}
               \hspace*{-0.01\textwidth}
               \includegraphics[width=0.54\textwidth,clip=]{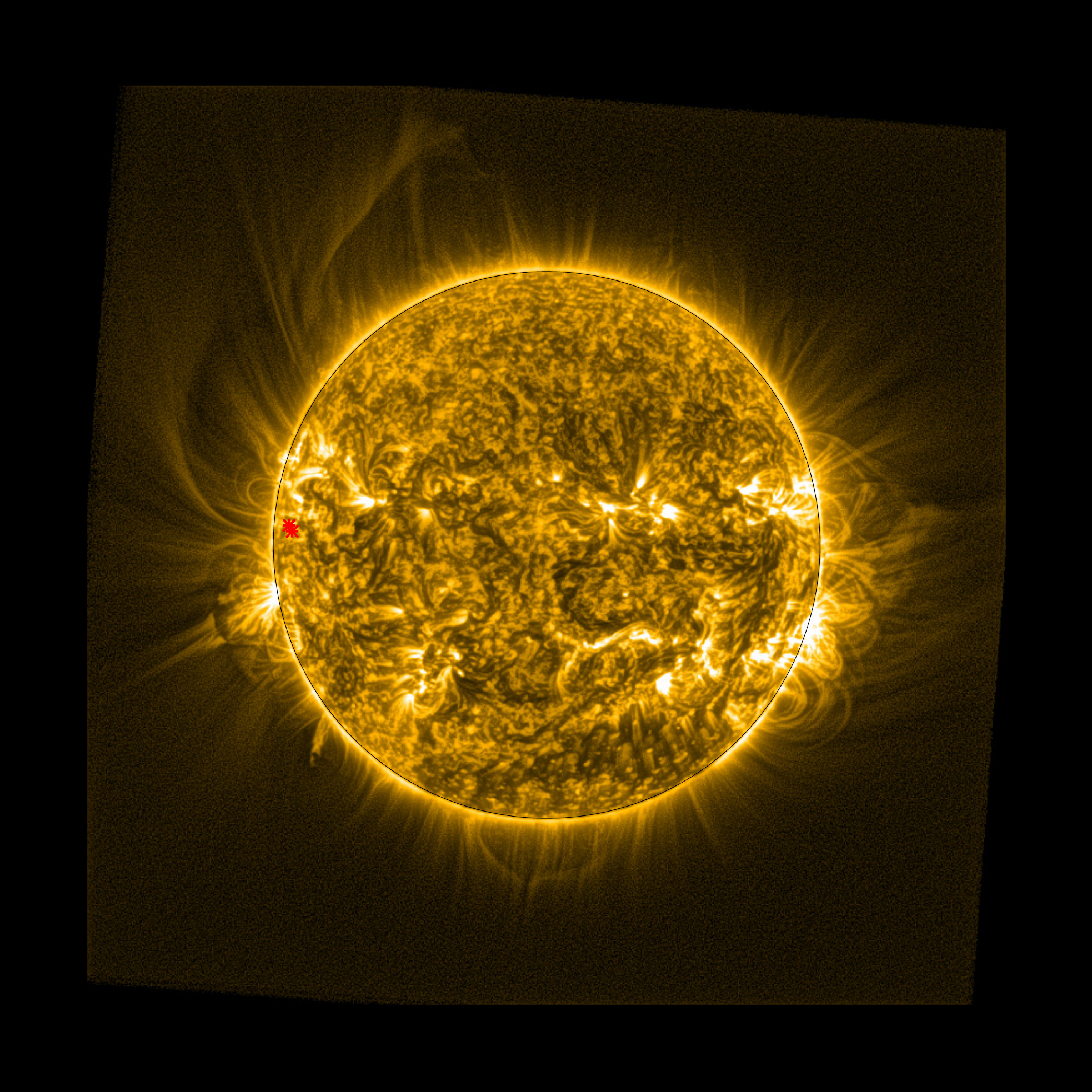}
              }
     \vspace{-0.04\textwidth}   % Shift close to the panel top 
     \centerline{\Large \bf     % Includes the labels (here needs the color 
                                %   package, see beginning of this file)
      \hspace{0.0\textwidth}  \color{black}\footnotesize{11-Oct-2014}
      \hspace{0.34\textwidth}  \color{white}{11-Oct-2014}
         \hfill}
     \vspace{0.02\textwidth}    % Shift back to the panel bottom 
     
   \centerline{\hspace*{0.015\textwidth}
               \includegraphics[width=0.54\textwidth,clip=]{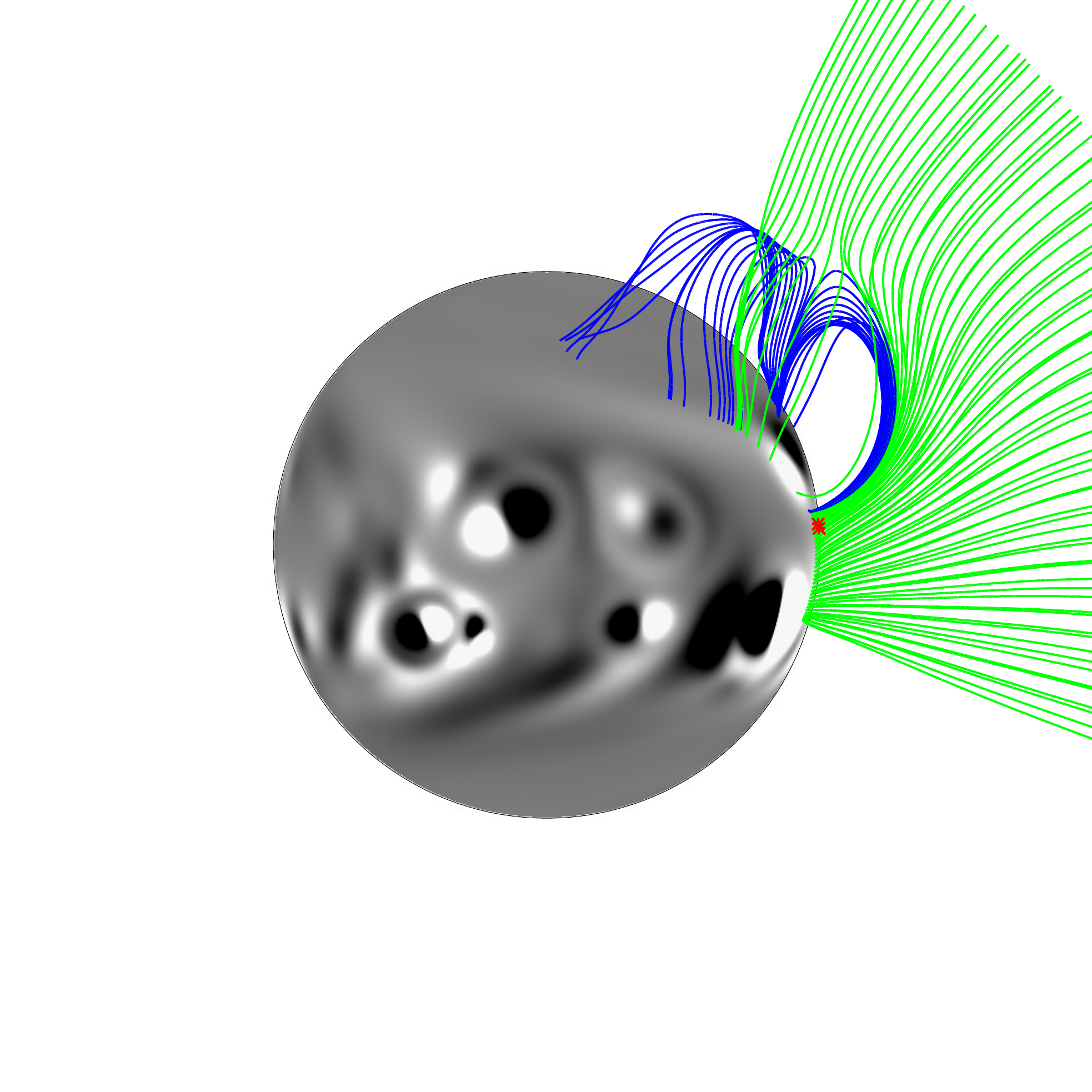}
               \hspace*{-0.01\textwidth}
               \includegraphics[width=0.54\textwidth,clip=]{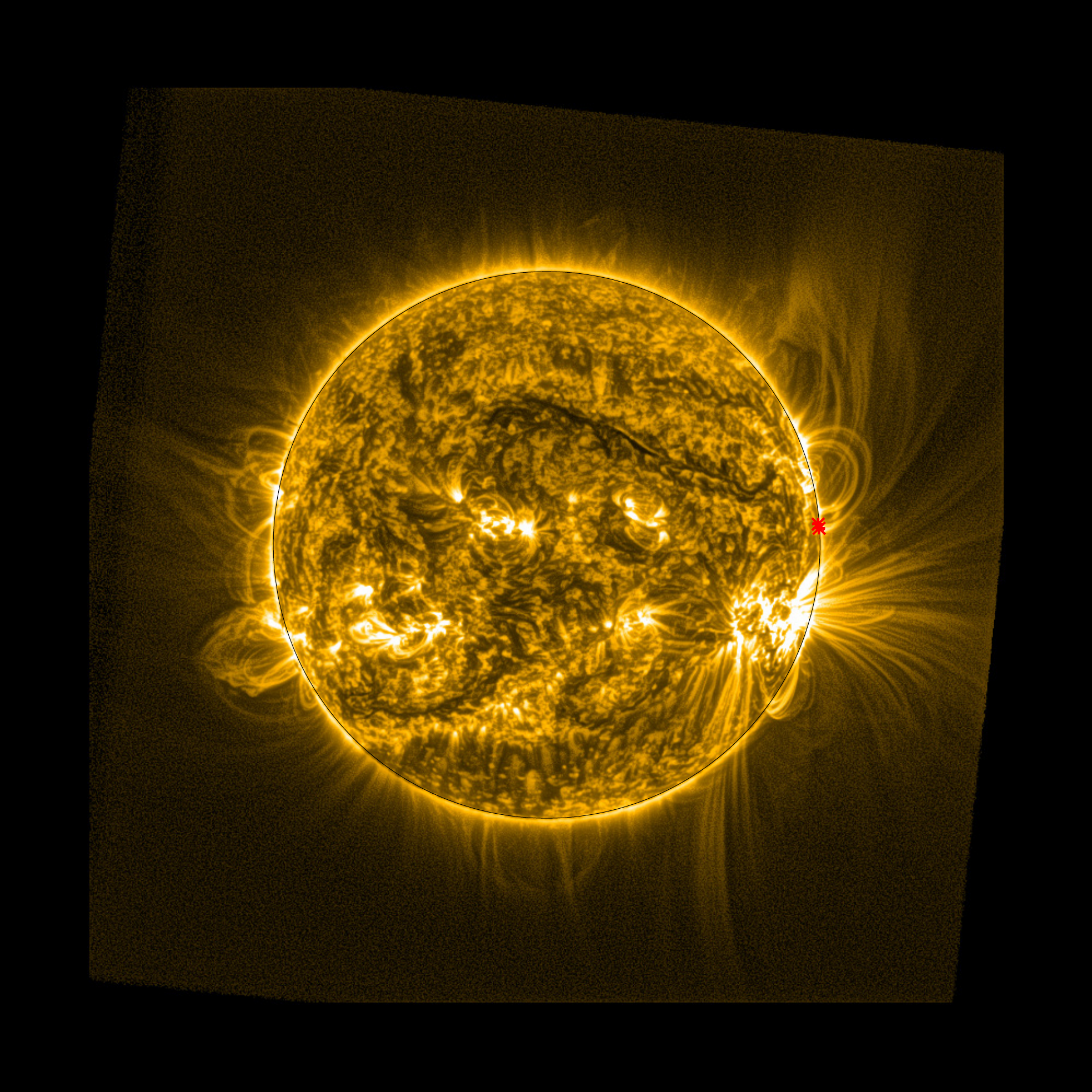}
              }
     \vspace{-0.04\textwidth}   % Shift close to the panel top 
     \centerline{\Large \bf     % Includes the labels (here needs the color 
                                %   package, see beginning of this file)
      \hspace{0.0\textwidth}  \color{black}\footnotesize{27-Oct-2014}
      \hspace{0.34\textwidth}  \color{white}{27-Oct-2014}
         \hfill}
     \vspace{0.02\textwidth}    % Shift back to the panel bottom 

\caption{Comparison between simulation coronal magnetic field and SWAP EUV observations of a persistent fan: Rotation 1. Top (11 October 2014), the fan can be seen off the NE limb of the Sun, bottom (27 October 2014), the fan can be seen off the NW limb. The approximate location of the fan footpoints are indicated by red stars {{(note that the two footpoints plotted are very close together here)}}. Simulation magnetic field lines were plotted by selecting starting points above the fan footpoints at a range of latitudes $\pm50$ degrees, at heights of 0.54\,$\rsun$ and 1.07\,$\rsun$ above the photosphere. Field lines are coloured dark blue (closed field), green (open, positive field), and light blue (open, negative field). An animation of this figure is included in the Electronic Supplementary Materials (fan1\_fl.mp4).}\label{fig:fan1a}
   \end{figure}

  \begin{figure}

   \centerline{\hspace*{0.015\textwidth}
               \includegraphics[width=0.54\textwidth,clip=]{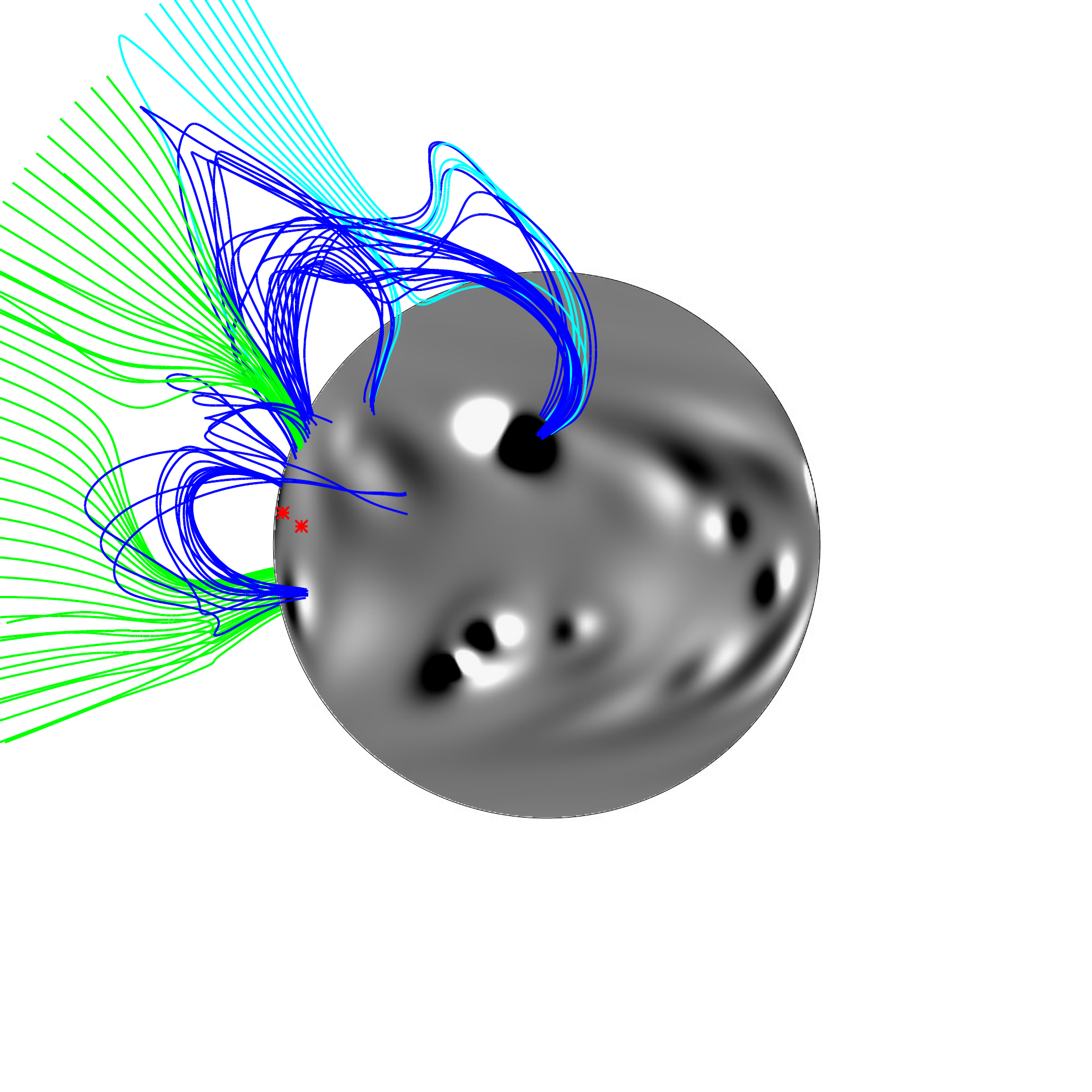}
               \hspace*{-0.01\textwidth}
               \includegraphics[width=0.54\textwidth,clip=]{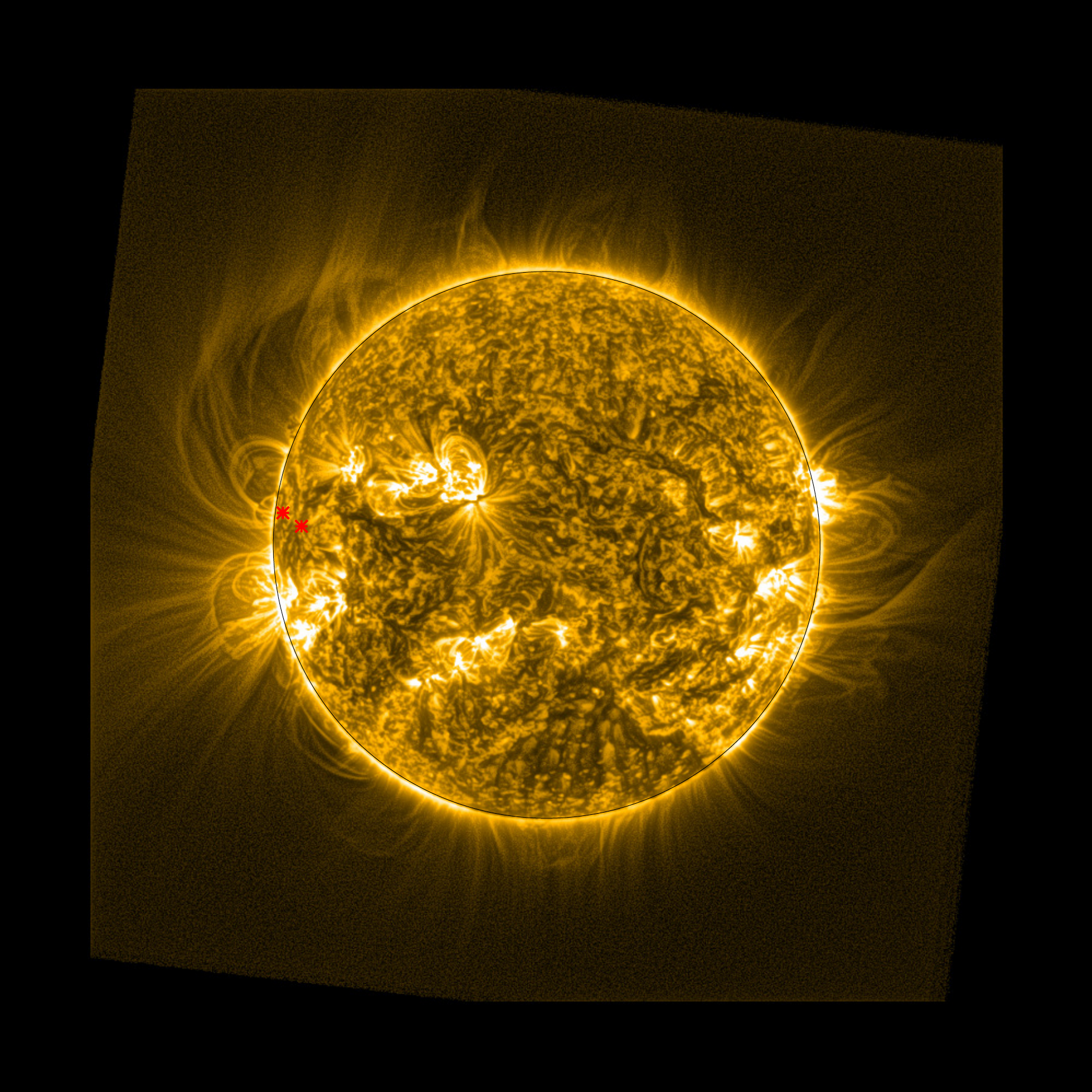}
              }
     \vspace{-0.04\textwidth}   % Shift close to the panel top 
     \centerline{\Large \bf     % Includes the labels (here needs the color 
                                %   package, see beginning of this file)
      \hspace{0.0\textwidth}  \color{black}\footnotesize{08-Nov-2014}
      \hspace{0.34\textwidth}  \color{white}{08-Nov-2014}
         \hfill}
     \vspace{0.02\textwidth}    % Shift back to the panel bottom 
     
   \centerline{\hspace*{0.015\textwidth}
               \includegraphics[width=0.54\textwidth,clip=]{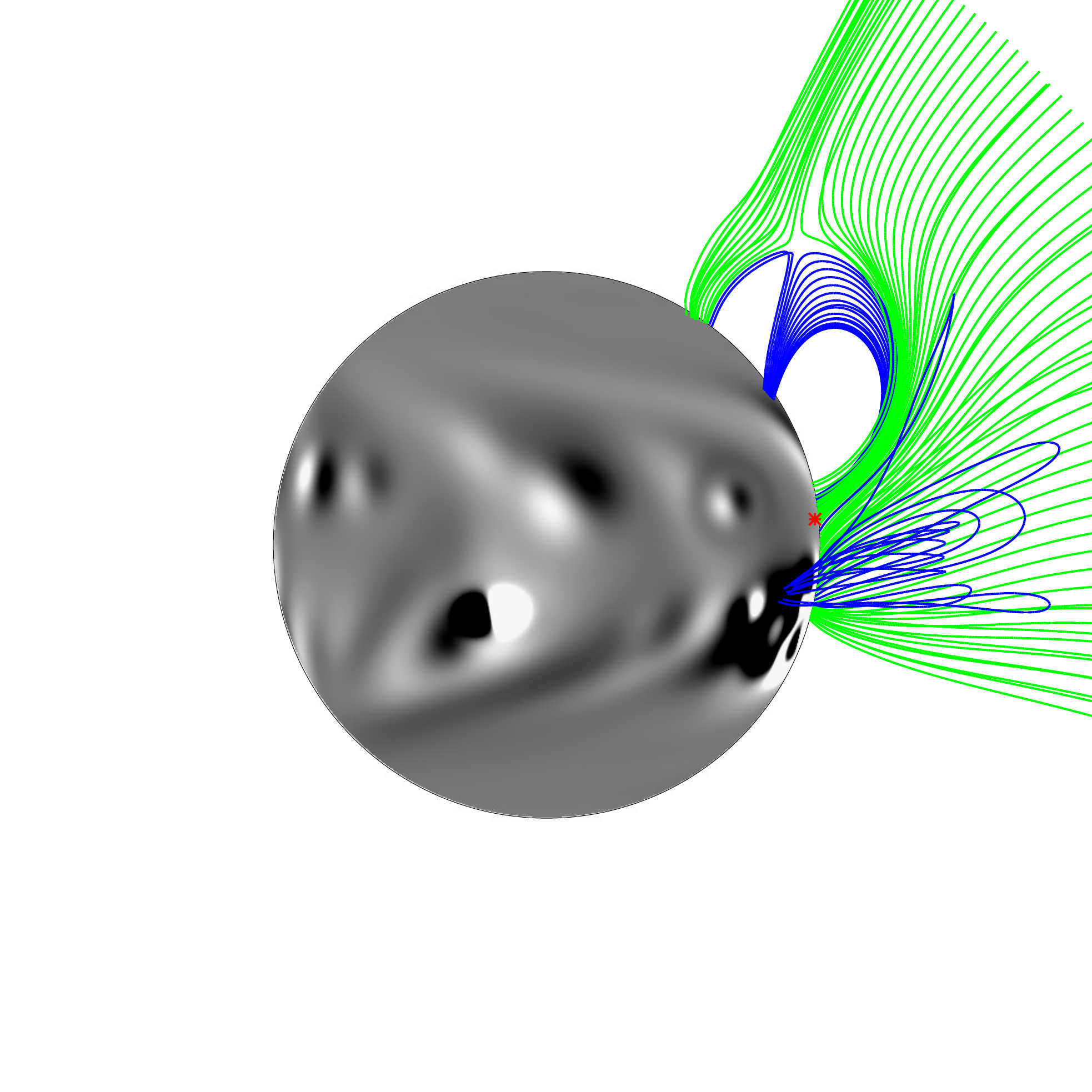}
               \hspace*{-0.01\textwidth}
               \includegraphics[width=0.54\textwidth,clip=]{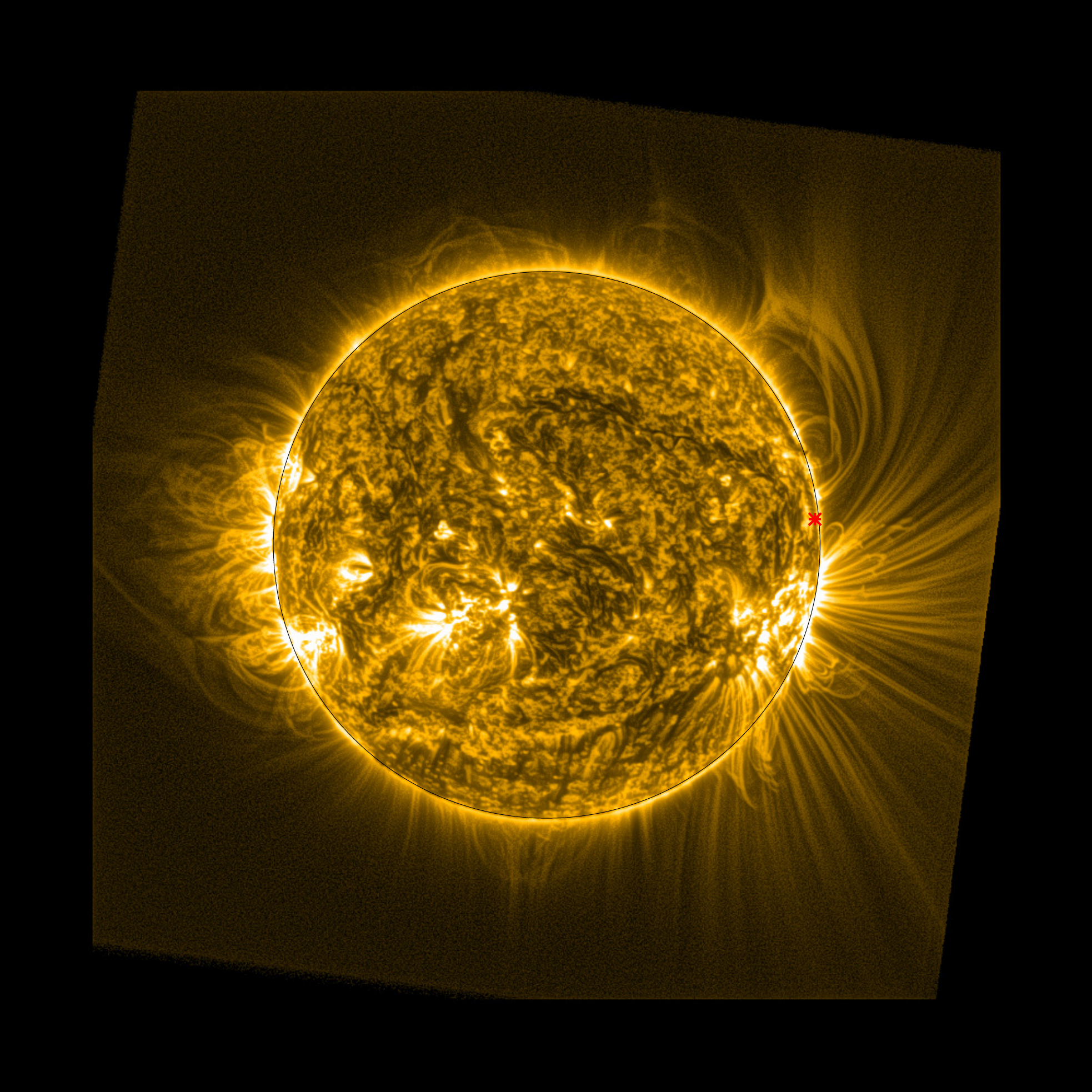}
              }
     \vspace{-0.04\textwidth}   % Shift close to the panel top 
     \centerline{\Large \bf     % Includes the labels (here needs the color 
                                %   package, see beginning of this file)
      \hspace{0.0\textwidth}  \color{black}\footnotesize{24-Nov-2014}
      \hspace{0.34\textwidth}  \color{white}{24-Nov-2014}
         \hfill}
     \vspace{0.02\textwidth}    % Shift back to the panel bottom 

\caption{Comparison between simulation coronal magnetic field and SWAP EUV observations of a persistent fan: Rotation 2. Top (8 November 2014), the fan can be seen off the NE limb of the Sun, bottom (24 November 2014), the fan can be seen off the NW limb. The approximate location of the fan footpoints are indicated by red stars. Simulation magnetic field lines were plotted by selecting starting points above the fan footpoints at a range of latitudes $\pm50$ degrees, at heights of 0.54\,$\rsun$ and 1.07\,$\rsun$ above the photosphere. Field lines are coloured dark blue (closed field), green (open, positive field), and light blue (open, negative field). An animation of this figure is included in the Electronic Supplementary Materials (fan1\_fl.mp4).}\label{fig:fan1b}
   \end{figure}

  \begin{figure}

   \centerline{\hspace*{0.015\textwidth}
               \includegraphics[width=0.54\textwidth,clip=]{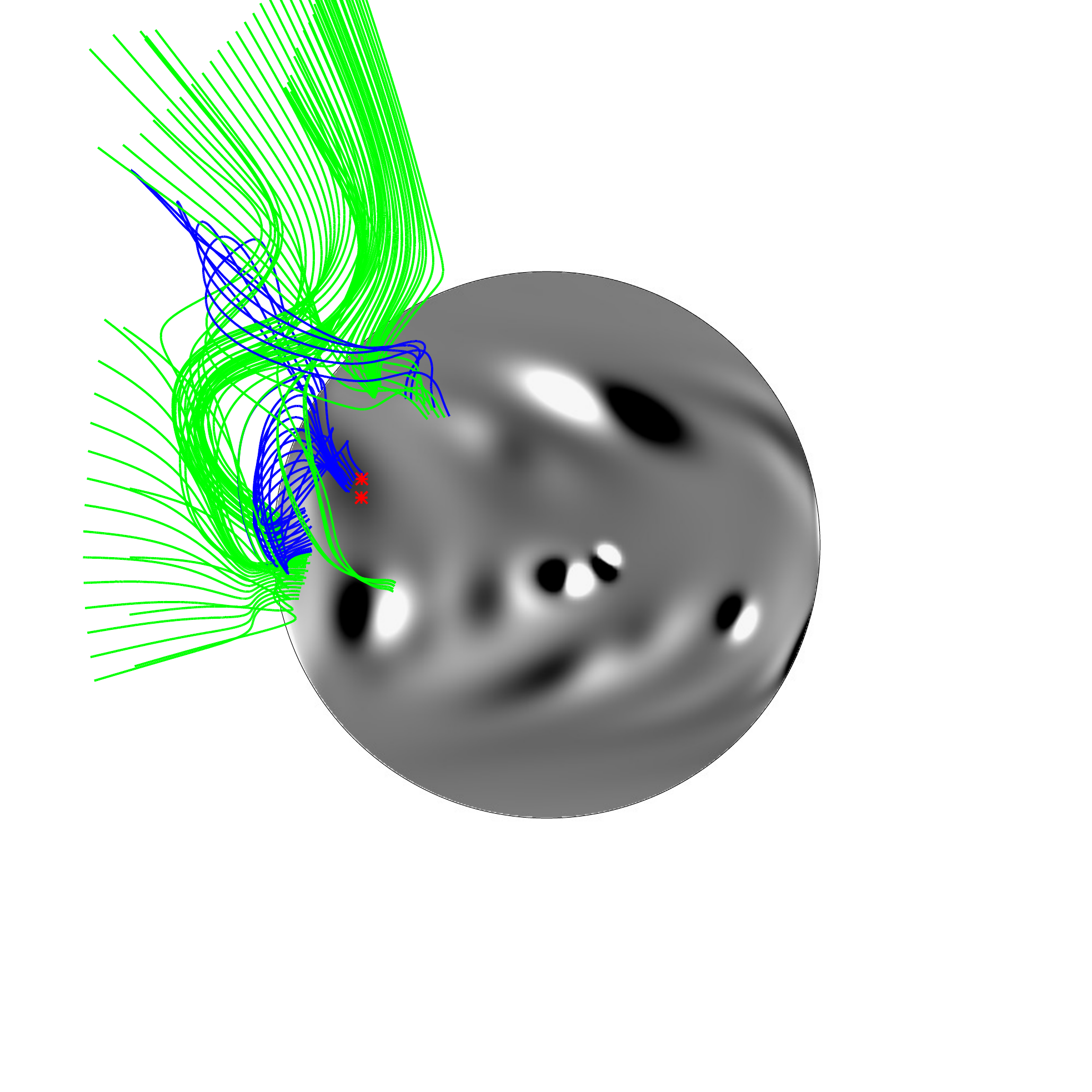}
               \hspace*{-0.01\textwidth}
               \includegraphics[width=0.54\textwidth,clip=]{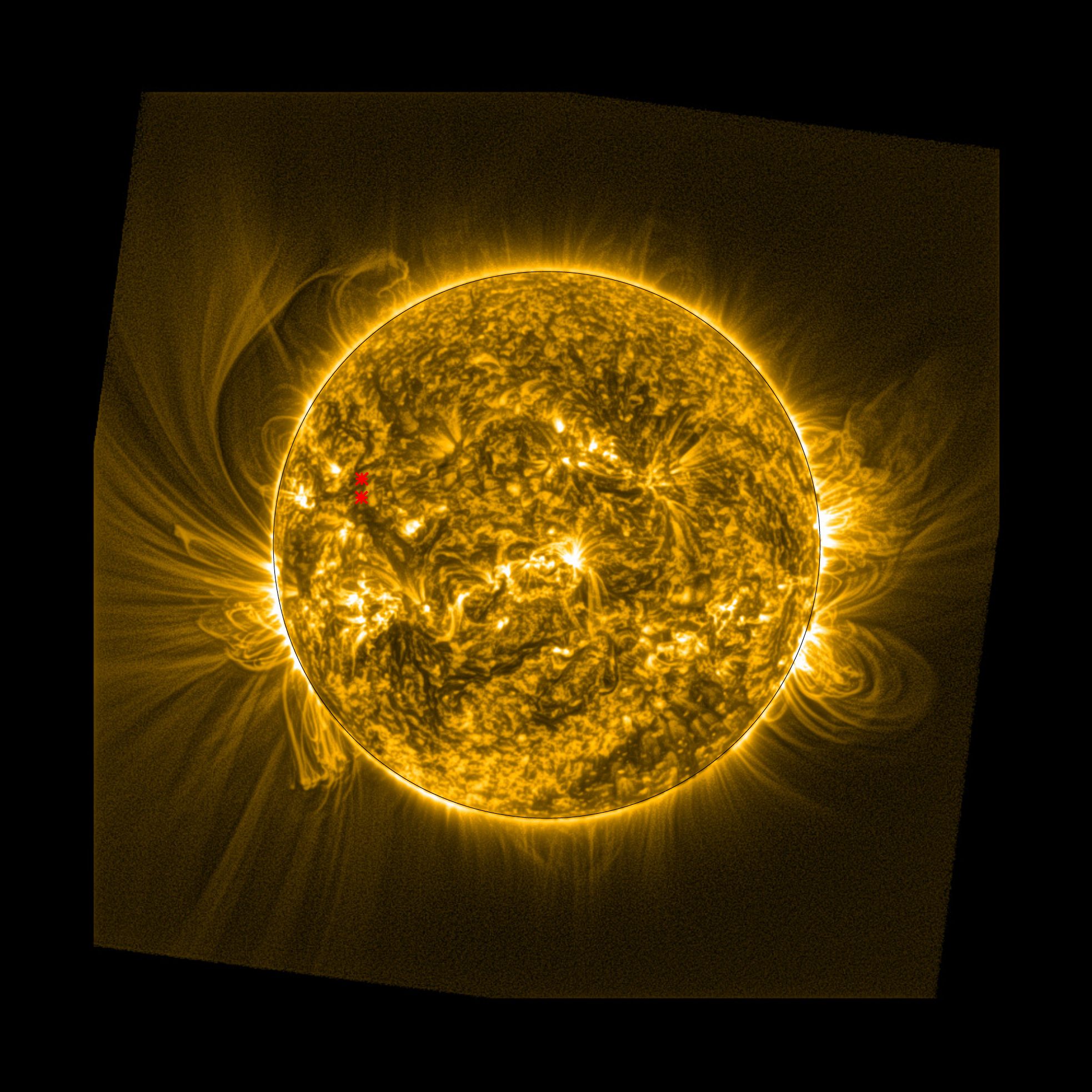}
              }
     \vspace{-0.04\textwidth}   % Shift close to the panel top 
     \centerline{\Large \bf     % Includes the labels (here needs the color 
                                %   package, see beginning of this file)
      \hspace{0.0\textwidth}  \color{black}\footnotesize{06-Dec-2014}
      \hspace{0.34\textwidth}  \color{white}{06-Dec-2014}
         \hfill}
     \vspace{0.02\textwidth}    % Shift back to the panel bottom 
     
   \centerline{\hspace*{0.015\textwidth}
               \includegraphics[width=0.54\textwidth,clip=]{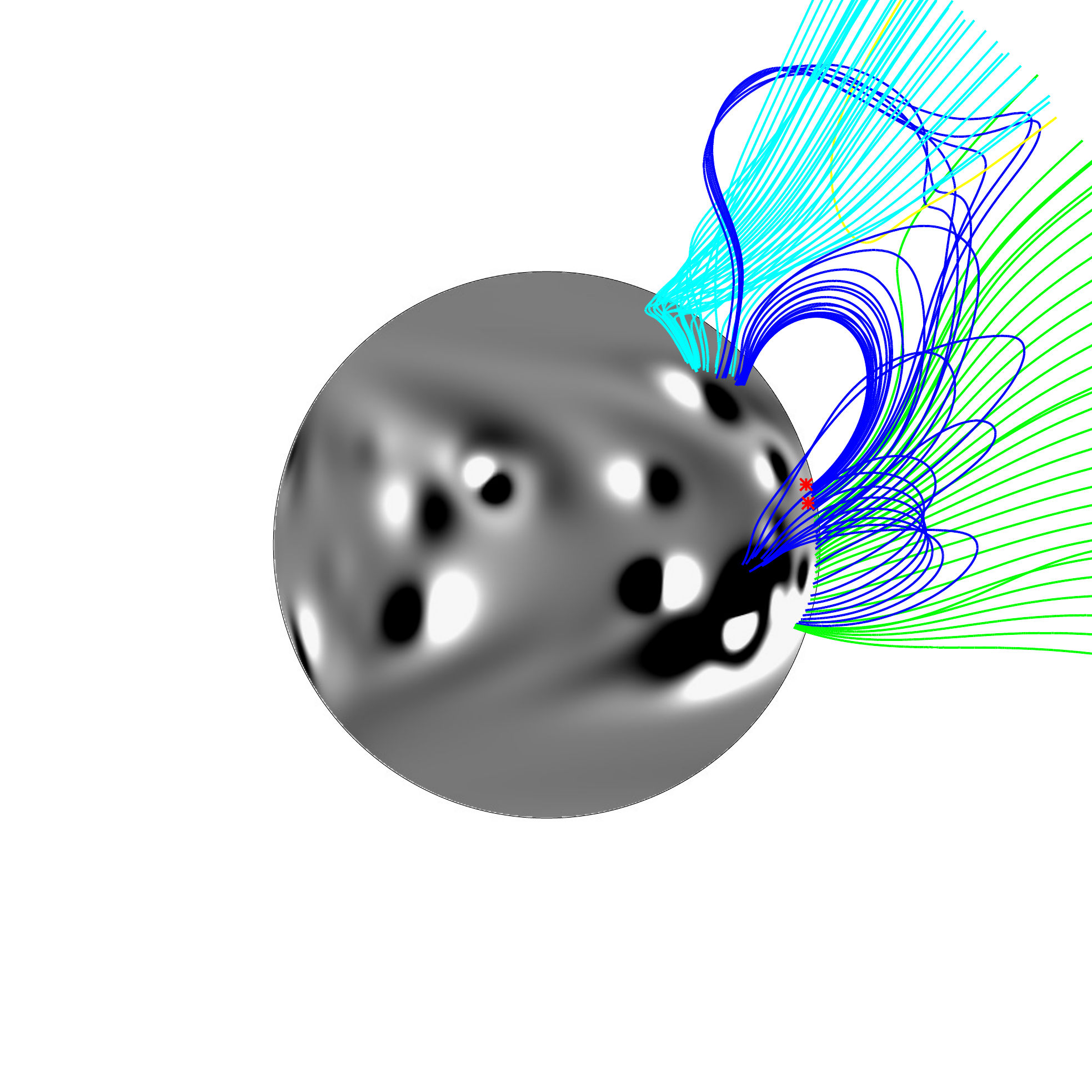}
               \hspace*{-0.01\textwidth}
               \includegraphics[width=0.54\textwidth,clip=]{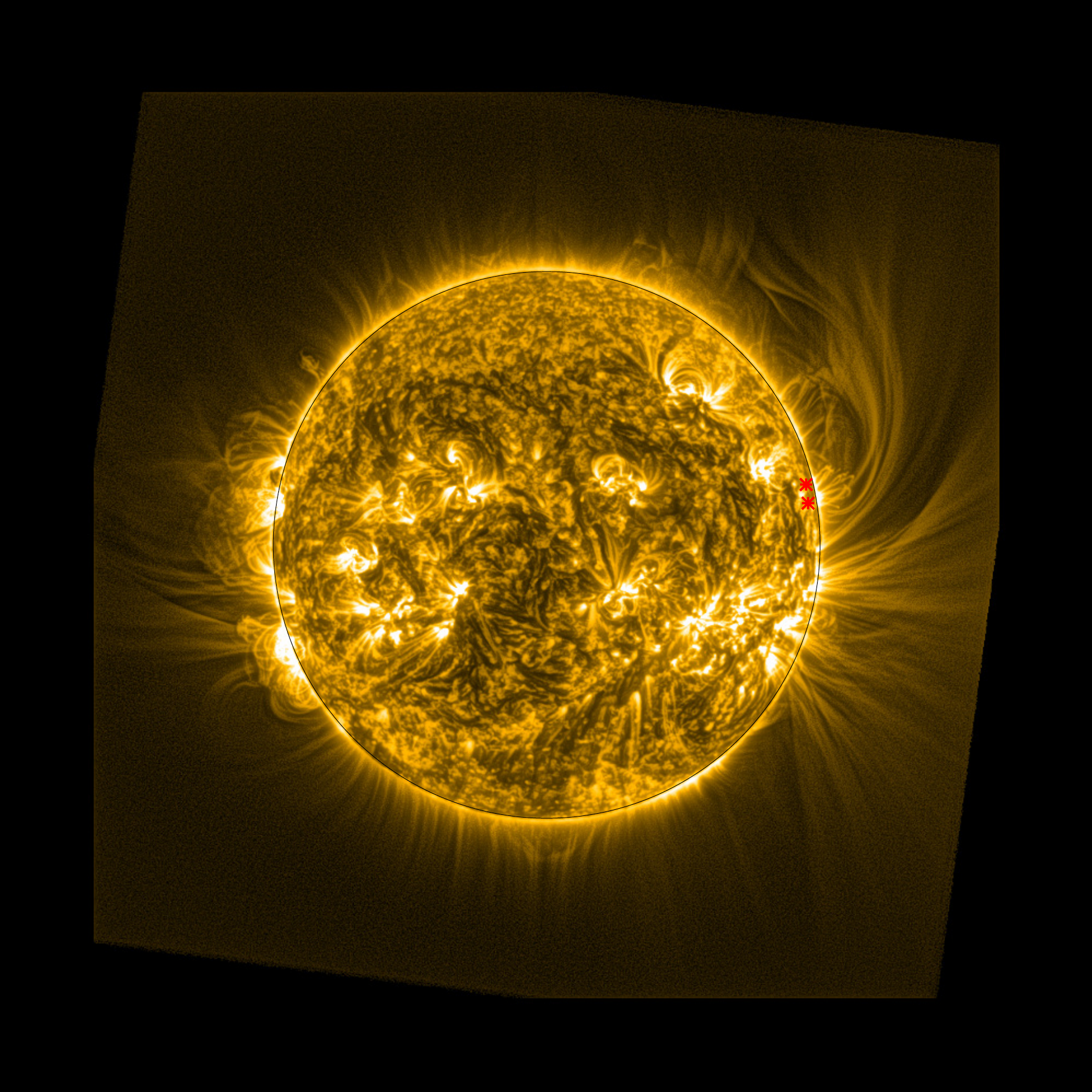}
              }
     \vspace{-0.04\textwidth}   % Shift close to the panel top 
     \centerline{\Large \bf     % Includes the labels (here needs the color 
                                %   package, see beginning of this file)
      \hspace{0.0\textwidth}  \color{black}\footnotesize{20-Dec-2014}
      \hspace{0.34\textwidth}  \color{white}{20-Dec-2014}
         \hfill}
     \vspace{0.02\textwidth}    % Shift back to the panel bottom 

\caption{Comparison between simulation coronal magnetic field and SWAP EUV observations of a persistent fan: Rotation 3. Top (6 December 2014), the fan can be seen off the NE limb of the Sun, bottom (20 December 2014), the fan can be seen off the NW limb. The approximate location of the fan footpoints are indicated by red stars. Simulation magnetic field lines were plotted by selecting starting points above the fan footpoints at a range of latitudes $\pm50$ degrees, at heights of 0.54\,$\rsun$ and 1.07\,$\rsun$ above the photosphere. Field lines are coloured dark blue (closed field), green (open, positive field), and light blue (open, negative field). An animation of this figure is included in the Electronic Supplementary Materials (fan1\_fl.mp4).}\label{fig:fan1c}
   \end{figure}

  \begin{figure}

   \centerline{\hspace*{0.015\textwidth}
               \includegraphics[width=0.54\textwidth,clip=]{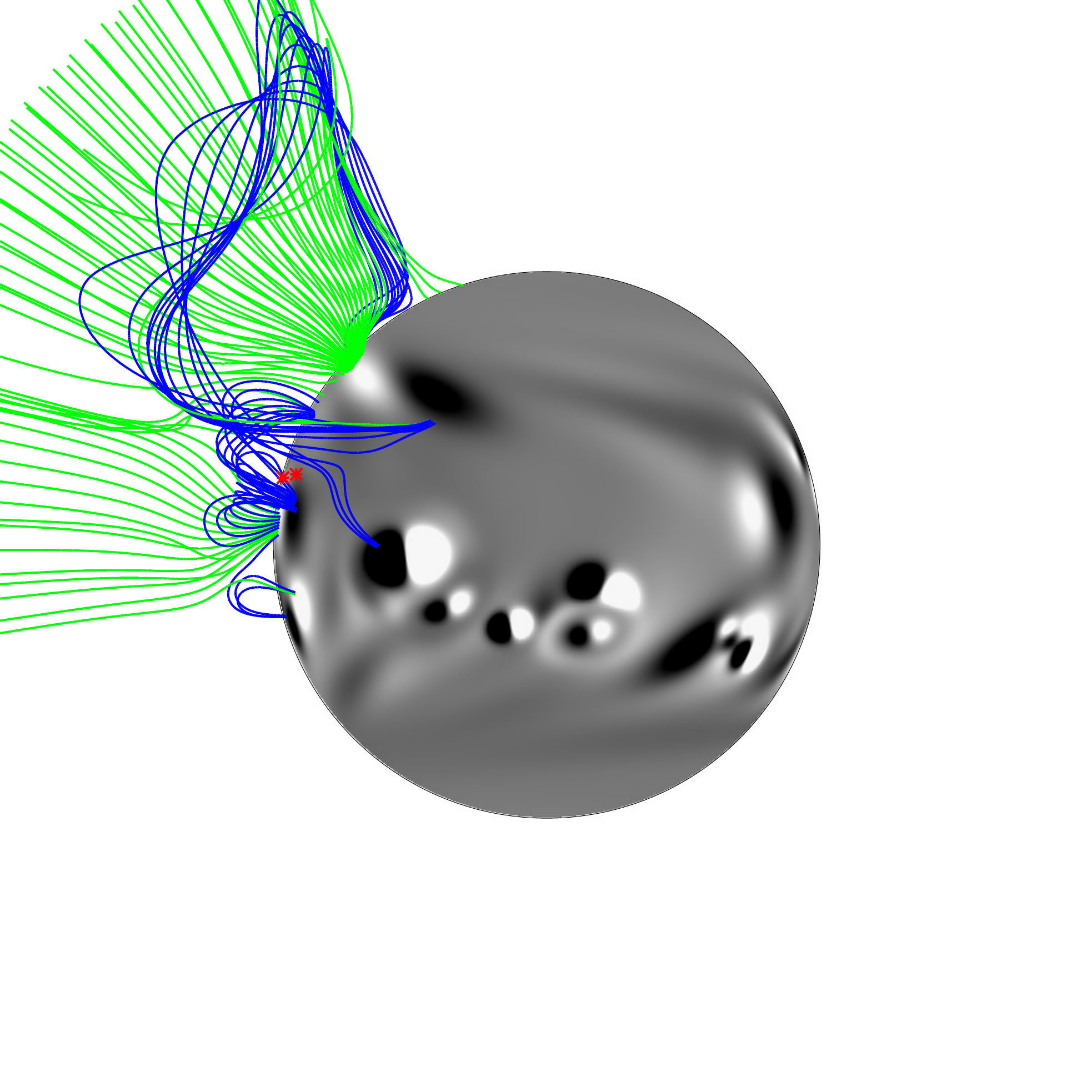}
               \hspace*{-0.01\textwidth}
               \includegraphics[width=0.54\textwidth,clip=]{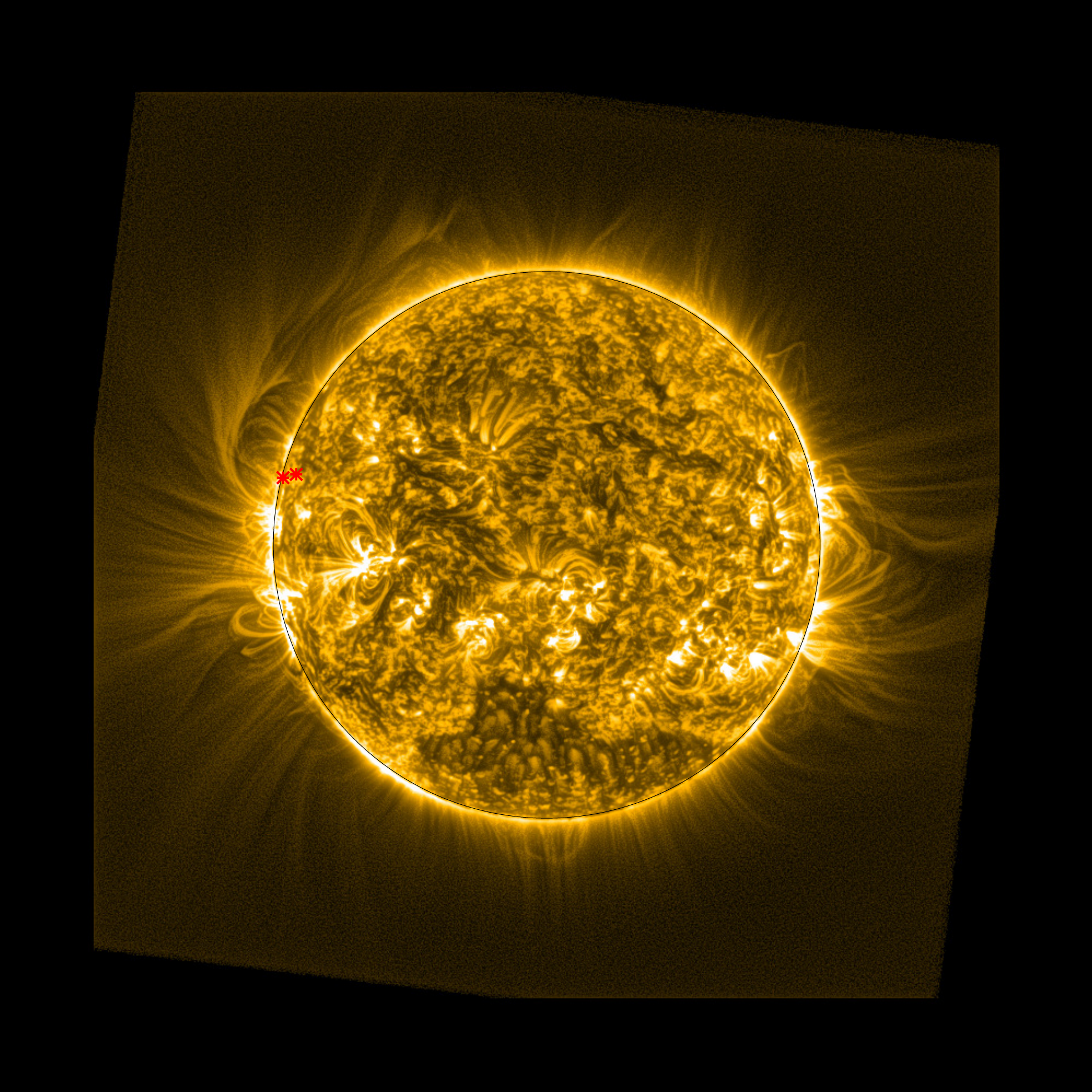}
              }
     \vspace{-0.04\textwidth}   % Shift close to the panel top 
     \centerline{\Large \bf     % Includes the labels (here needs the color 
                                %   package, see beginning of this file)
      \hspace{0.0\textwidth}  \color{black}\footnotesize{01-Jan-2015}
      \hspace{0.34\textwidth}  \color{white}{01-Jan-2015}
         \hfill}
     \vspace{0.02\textwidth}    % Shift back to the panel bottom 
     
   \centerline{\hspace*{0.015\textwidth}
               \includegraphics[width=0.54\textwidth,clip=]{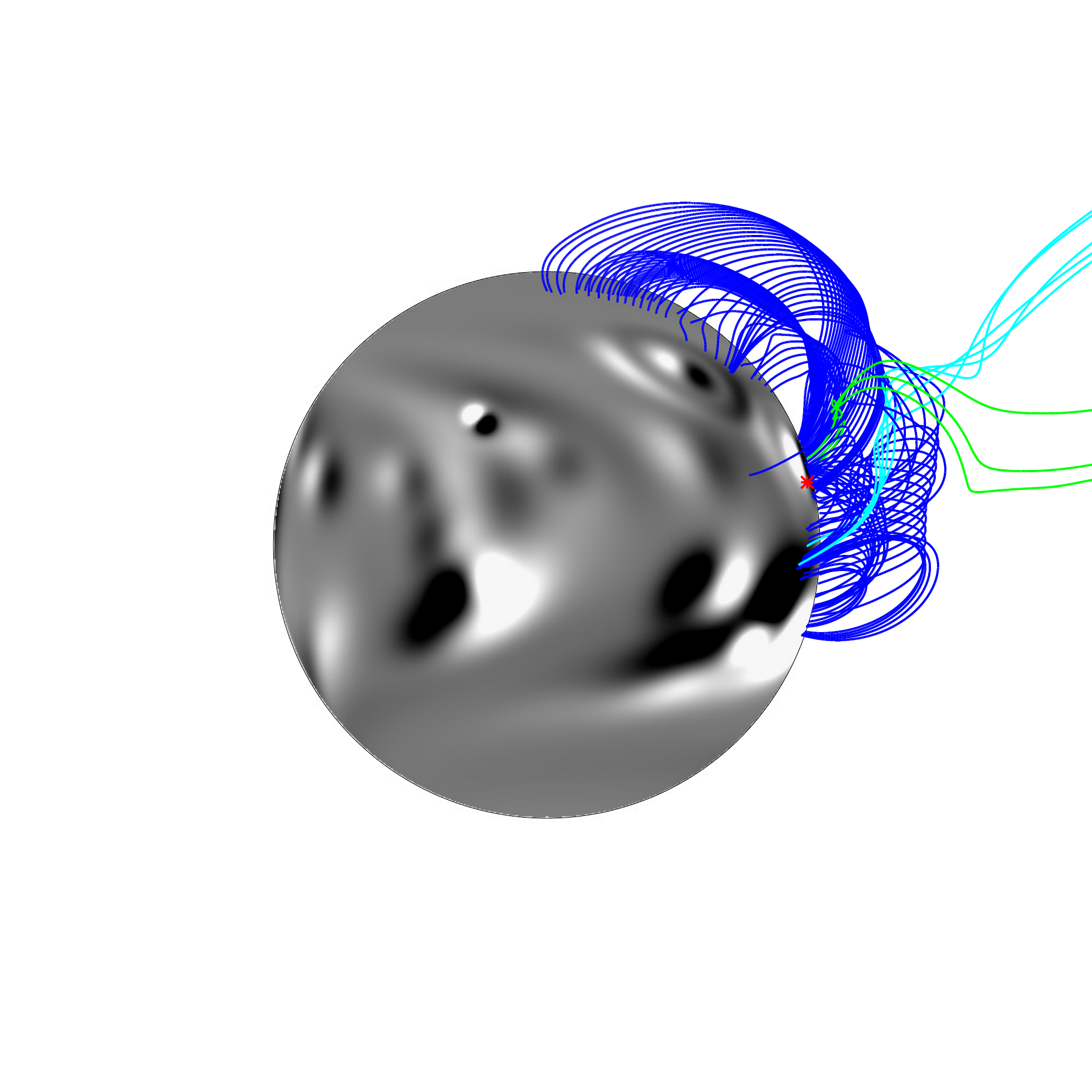}
               \hspace*{-0.01\textwidth}
               \includegraphics[width=0.54\textwidth,clip=]{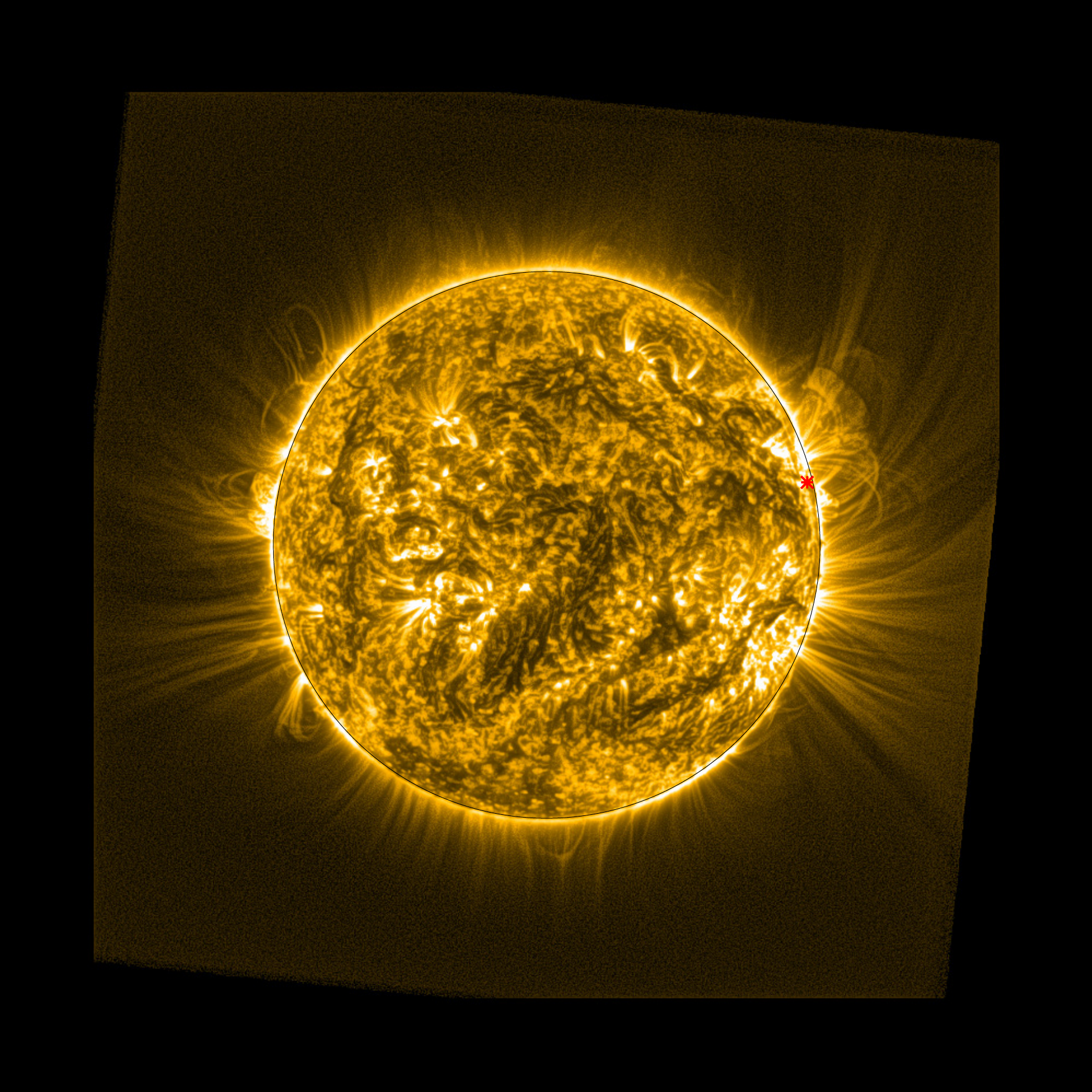}
              }
     \vspace{-0.04\textwidth}   % Shift close to the panel top 
     \centerline{\Large \bf     % Includes the labels (here needs the color 
                                %   package, see beginning of this file)
      \hspace{0.0\textwidth}  \color{black}\footnotesize{15-Jan-2015}
      \hspace{0.34\textwidth}  \color{white}{15-Jan-2015}
         \hfill}
     \vspace{0.02\textwidth}    % Shift back to the panel bottom 

\caption{Comparison between simulation coronal magnetic field and SWAP EUV observations of a persistent fan: Rotation 4. Top (1 January 2015), the fan can be seen off the NE limb of the Sun, bottom (15 January 2015), the fan can be seen off the NW limb. The approximate location of the fan footpoints are indicated by red stars. Simulation magnetic field lines were plotted by selecting starting points above the fan footpoints at a range of latitudes $\pm50$ degrees, at heights of 0.27\,$\rsun$ and 0.54\,$\rsun$ above the photosphere (note that for the 4th rotation, lower heights were used for starting points than in rotations 1\,--\,3, as the fan has significantly reduced in size). Field lines are coloured dark blue (closed field), green (open, positive field), and light blue (open, negative field). An animation of this figure is included in the Electronic Supplementary Materials (fan1\_fl.mp4).}\label{fig:fan1d}
   \end{figure}

In this section we discuss the evolution of a coronal fan that was observed off the north-east limb in early October, and persisted for five Carrington rotations. Coronal fans appear to form at the interface between regions of closed and open magnetic field, often overlying cusp-shaped voids \citep{seaton2013}. Here, we compare large-scale, off-limb structures pre and post disc transit to determine how well they are reproduced by the simulation. Figures~\ref{fig:fan1a}\,--\,\ref{fig:fan1d} show SWAP EUV observations of the fan when it was visible on the north-east and north-west solar limbs during the first four rotations, side-by-side with images from the simulation on the same dates.
  \begin{figure}

   \centerline{\hspace*{0.015\textwidth}
               \includegraphics[width=0.54\textwidth,clip=]{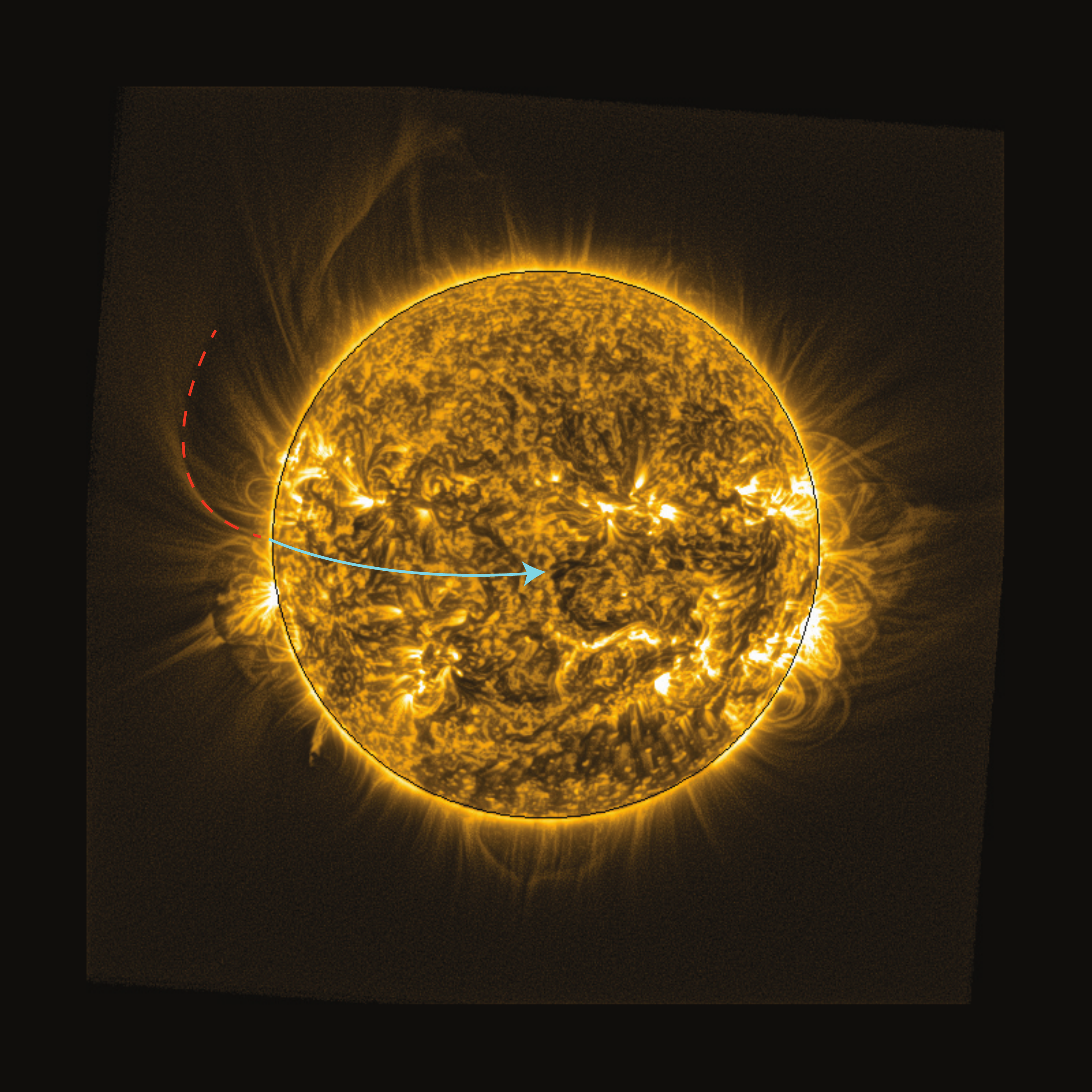}
               \hspace*{-0.01\textwidth}
               \includegraphics[width=0.54\textwidth,clip=]{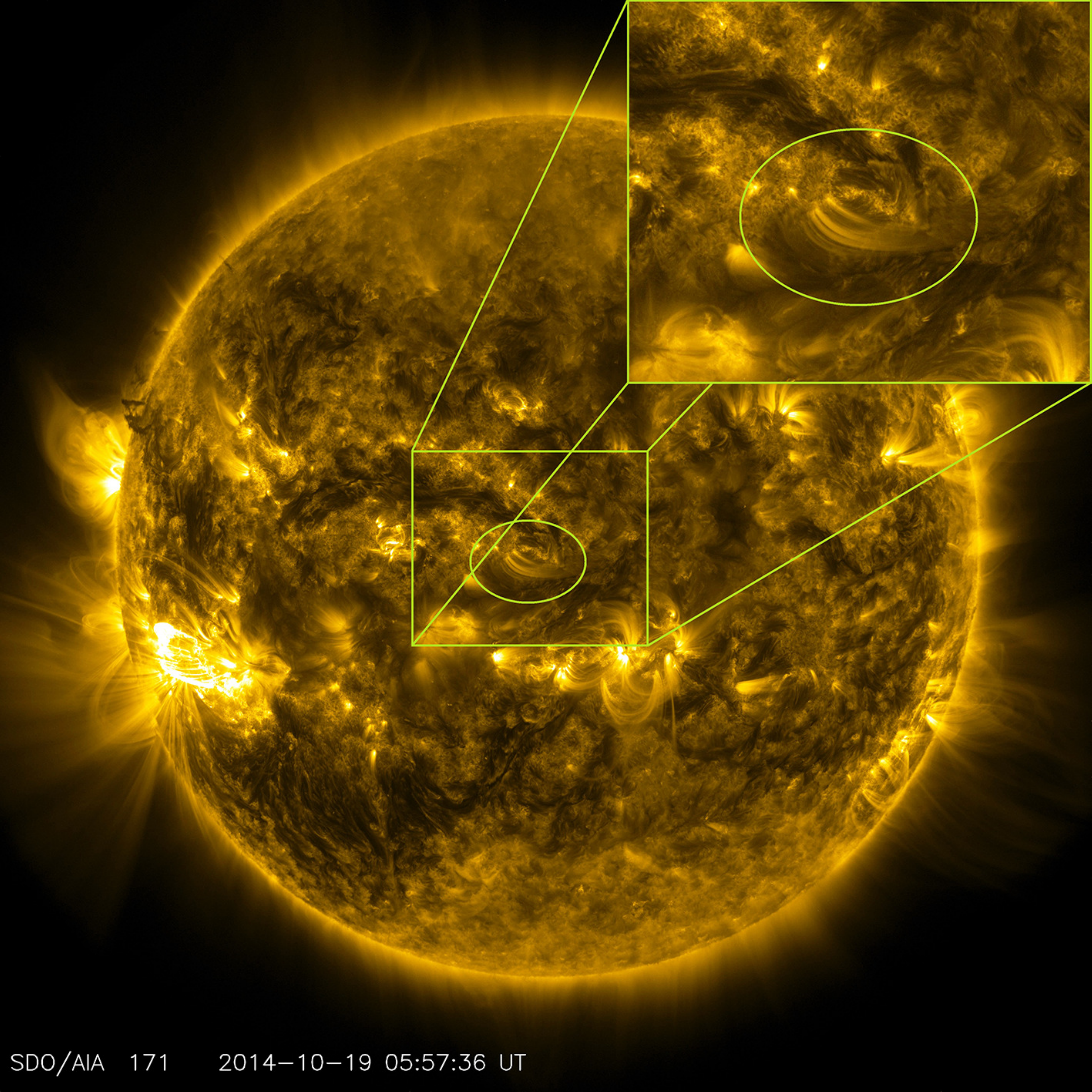}
              }
     \vspace{-0.54\textwidth}   % Shift close to the panel top 
     \centerline{\Large \bf     % Includes the labels (here needs the color 
                                %   package, see beginning of this file)
      \hspace{-0.04\textwidth}  \color{white}\small{(a)}
      \hspace{0.49\textwidth}  \color{white}\small{(b)}
         \hfill}
     \vspace{0.52\textwidth}    % Shift back to the panel bottom 
     \vspace{-0.08\textwidth}   % Shift close to the panel top 
     \centerline{\Large \bf     % Includes the labels (here needs the color 
                                %   package, see beginning of this file)
      \hspace{-0.02\textwidth}  \color{white}\footnotesize{11-Oct-2014}
         \hfill}
     \vspace{0.04\textwidth} 
\caption{{{(a) SWAP EUV image from 11 October 2014: illustration of how the fan structure is traced to the surface when it is observed off the east limb (red dashed line), in order to identify its footpoint locations. The footpoints are then tracked as the Sun rotates (blue line). (b) AIA 171\,\AA\, image from 19 October 2014. The ellipse indicates the location of the fan footpoints when they were located near central meridian, seen as a pair of small bright arcs.}}}\label{fig:fanfoot}
   \end{figure}
   
Fan footpoints are typically associated with brights points observed in EUV on-disc images \citep{talpeanu2016}.
{{The approximate location of the fan footpoints were determined by tracing the fan structure to bright points on the solar surface, while the fan was visible on the east limb (illustrated by the red dashed line in Fig.~\ref{fig:fanfoot}a, which is overplotted on a SWAP EUV image from 11 October 2014). The footpoints were then tracked as the Sun rotated (illustrated by the blue line in Fig.~\ref{fig:fanfoot}a), and their coordinates extracted when they had rotated approximately to disc centre. Figure~\ref{fig:fanfoot}b shows an AIA 171\,\AA\, image from 19 October 2014, with an ellipse to indicate the footpoint locations. These can be seen as two small, bright arcs. Note that while the footpoints can be tracked accurately when visible on the Earth-facing side of the Sun, it is not currently possible to determine their exact locations after they have rotated to the far side, since they are not necessarily static and will move due to underlying plasma motions. Therefore, the footpoint locations are re-determined each rotation, when the fan is visible on the east limb.}}

\begin{figure}
   \centerline{\hspace*{0.0\textwidth}
               \includegraphics[width=0.5\textwidth,clip=]{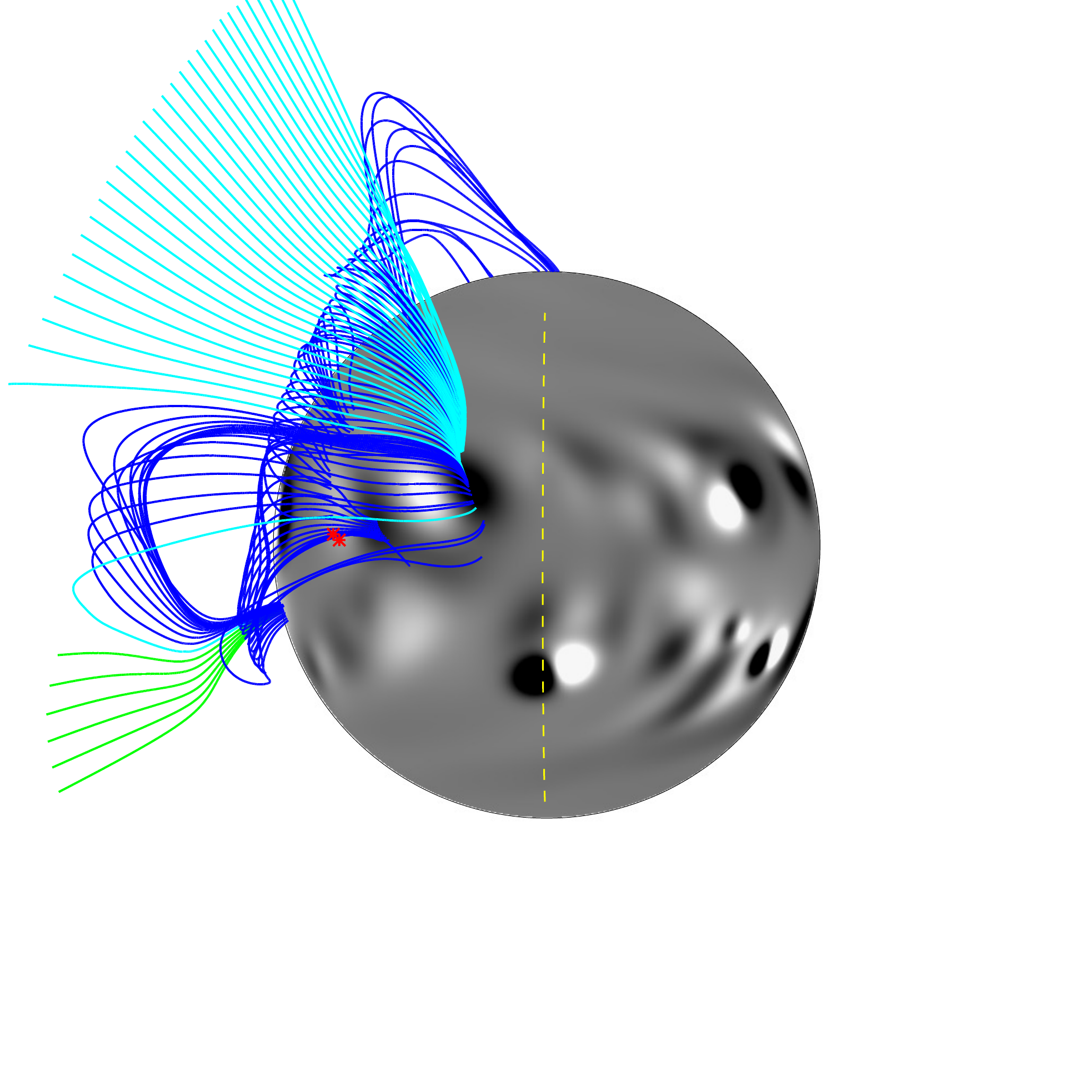}
               \hspace*{-0.0\textwidth}
               \includegraphics[width=0.5\textwidth,clip=]{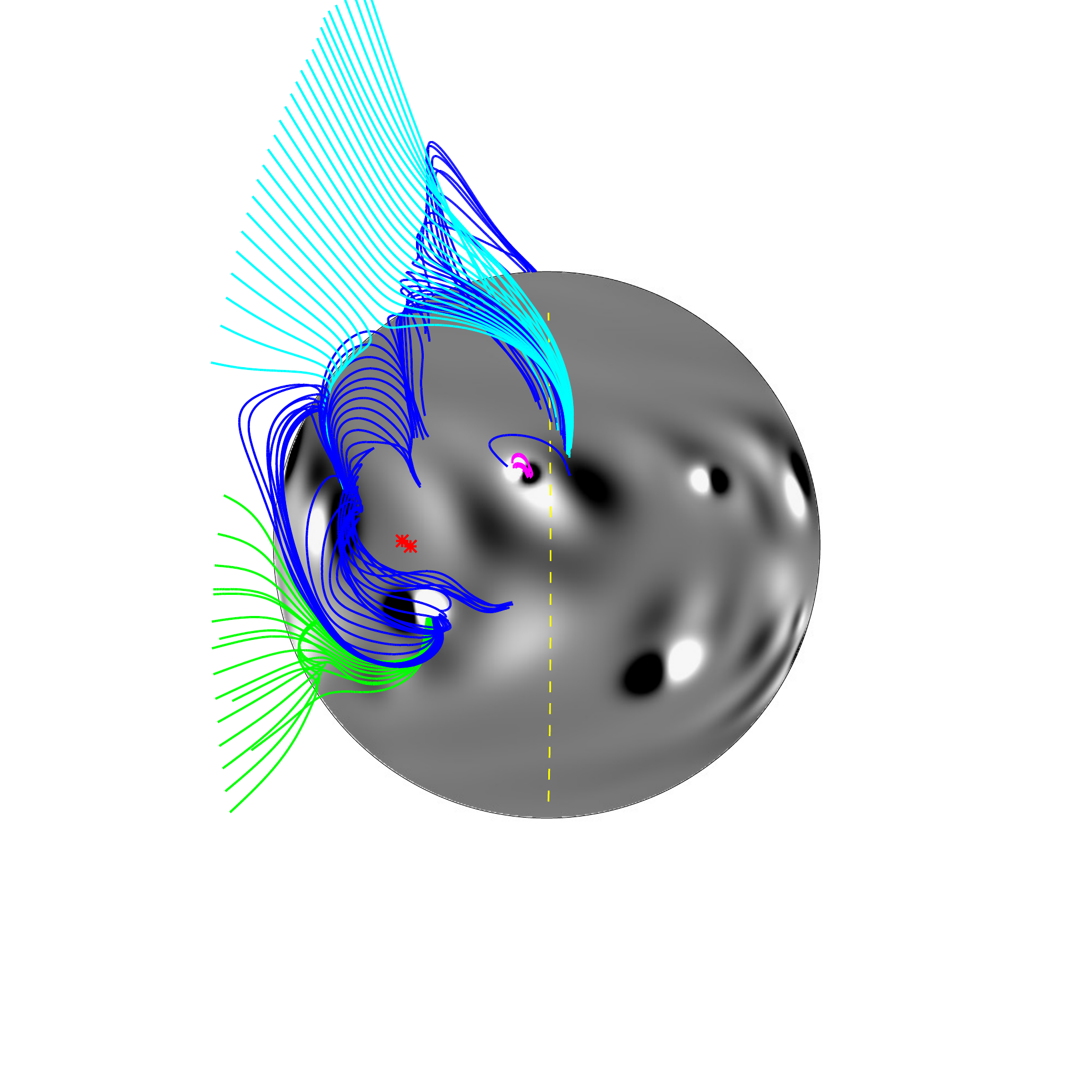}
              }
     \vspace{-0.5\textwidth}   % Shift close to the panel top 
     \centerline{\Large \bf     % Includes the labels (here needs the color 
                                %   package, see beginning of this file)
      \hspace{-0.1 \textwidth}   \color{black}\footnotesize{13-Oct-2014}
      \hspace{0.4\textwidth}  \color{black}{15-Oct-2014}
         \hfill}
     \vspace{0.46\textwidth}    % Shift back to the panel bottom 

   \centerline{\hspace*{0.0\textwidth}
               \includegraphics[width=0.5\textwidth,clip=]{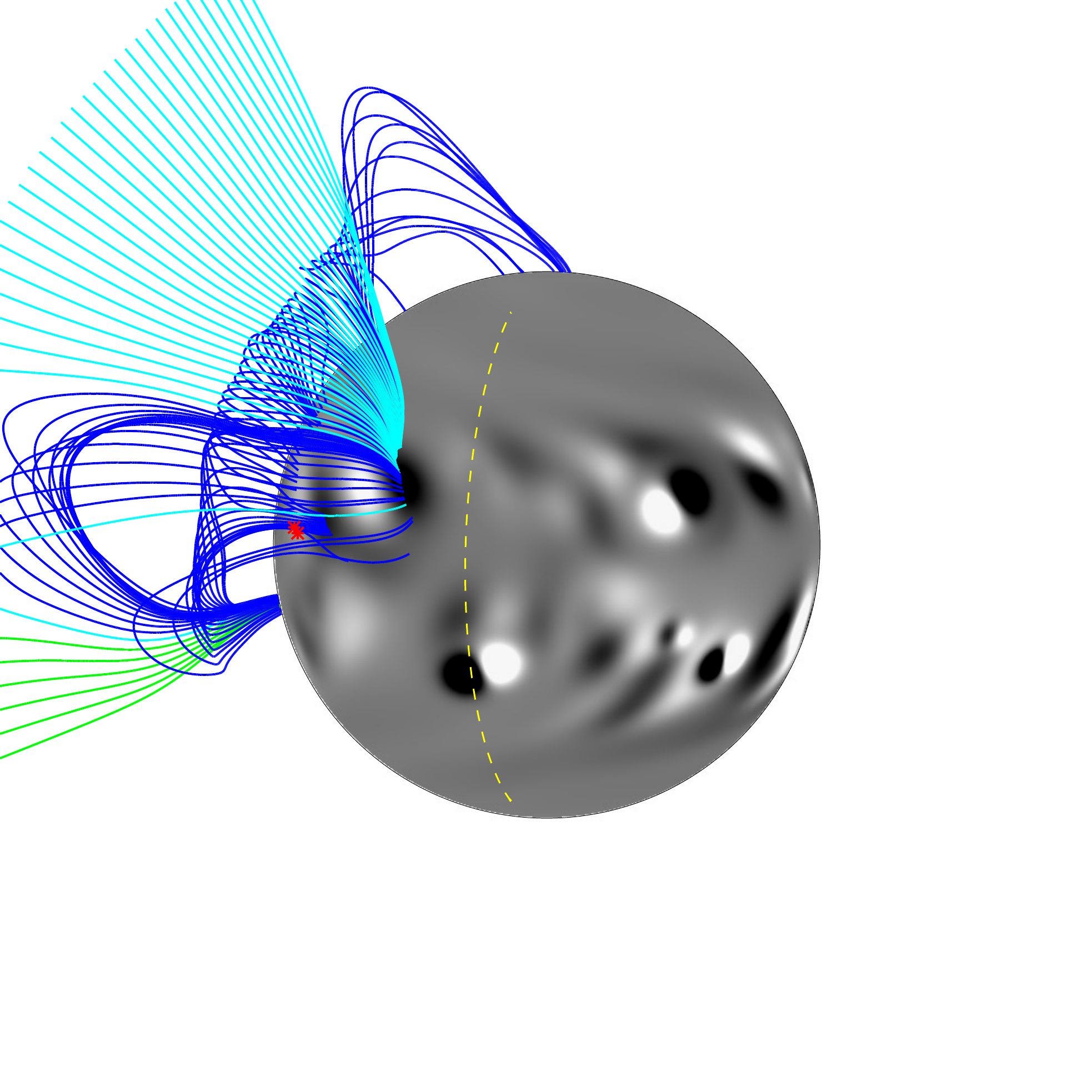}
               \hspace*{-0.0\textwidth}
               \includegraphics[width=0.5\textwidth,clip=]{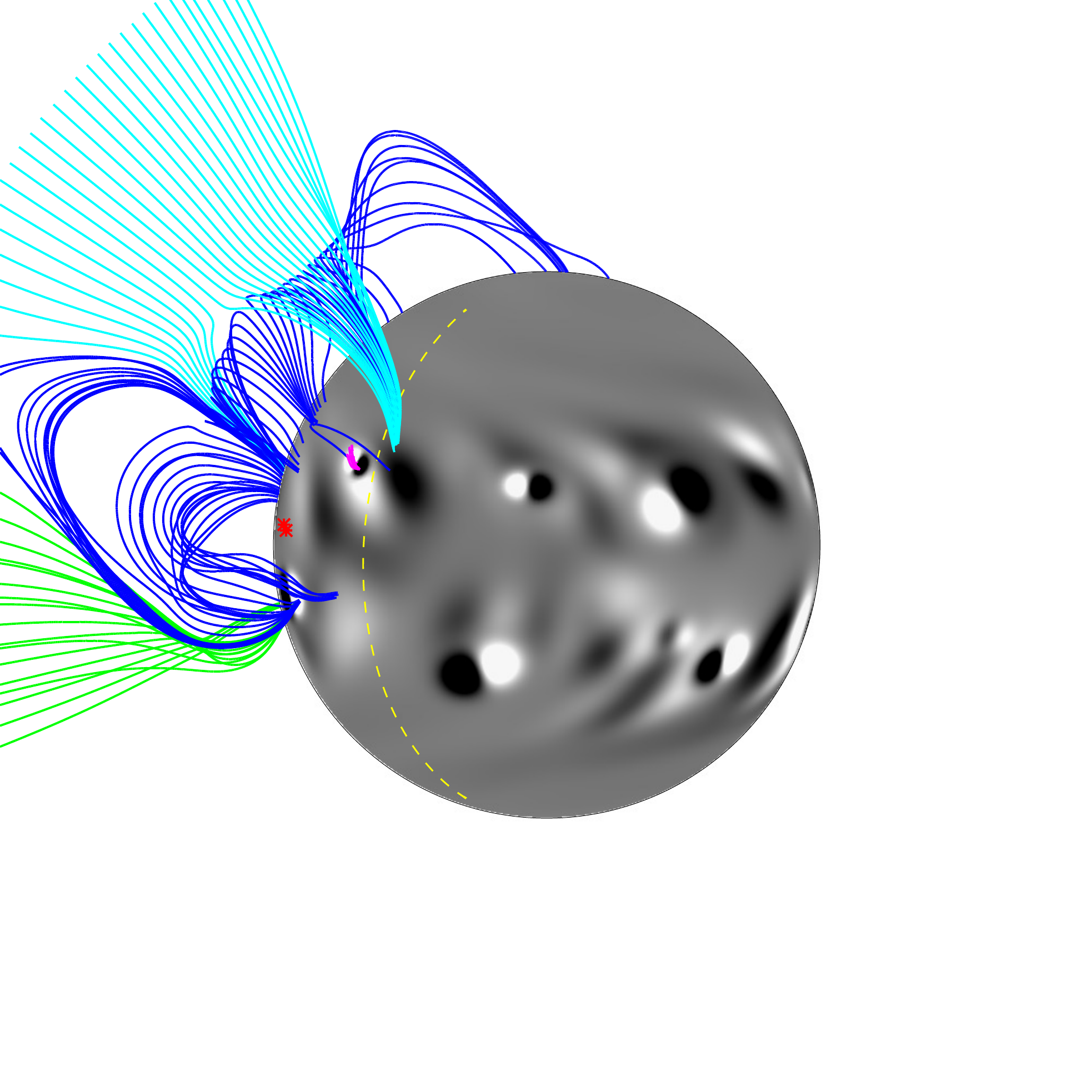}
              }
     \vspace{-0.5\textwidth}   % Shift close to the panel top 
     \centerline{\Large \bf     % Includes the labels (here needs the color 
                                %   package, see beginning of this file)
      \hspace{-0.1 \textwidth}   \color{black}\footnotesize{13-Oct-2014}
      \hspace{0.4\textwidth}  \color{black}{15-Oct-2014}
         \hfill}
     \vspace{0.46\textwidth}    % Shift back to the panel bottom 

\caption{Simulated magnetic field, with field lines plotted above the fan starting points on 13 and 15 October. Field lines are coloured and starting points chosen in the same manner as described in Fig.~\ref{fig:fan1a}. The approximate location of the fan footpoints are indicated by red stars {{(note that the two footpoints plotted are very close together here)}}. The top row shows the Earth-facing view of the Sun. In the bottom row, the Sun has been rotated so that the fan footpoints are close to the limb. In all images, central meridian is indicated by the yellow dashed line.}\label{fig:fan2}
   \end{figure}

{{For the purpose of overplotting magnetic field lines in the simulation, the footpoint locations from central meridian were translated forward and backward in time using the solar rotation rate and differential rotation profile (Eq.~\ref{eqn:DR}). This was used to estimate their positions as they crossed the Earth-facing side of the Sun.
These estimated footpoints are overplotted as red stars on both the EUV and simulation images in Figs.~\ref{fig:fan1a}\,--\,\ref{fig:fan1d} and \ref{fig:fan2}.}}
Magnetic field lines from the simulation were plotted by selecting starting points above the fan footpoints at a range of latitudes $\pm50$ degrees, at heights of 0.54\,$\rsun$ and 1.07\,$\rsun$ above the photosphere. These are the same heights as the arcs indicated in Fig.~\ref{fig:056} (note that structures closing below 0.54\,$\rsun$ are not captured here due to this). Closed field lines are coloured dark blue, positive polarity open field lines are coloured green, and negative polarity open field lines are coloured light blue. For 1 and 15 January (Fig.~\ref{fig:fan1d}), heights of 0.27\,$\rsun$ and 0.54\,$\rsun$ were used for starting points instead, as the fan has significantly reduced in size.

An animation of Figs.~\ref{fig:fan1a}\,--\,\ref{fig:fan1d} is included in the Electronic Supplementary Materials (fan1\_fl.mp4). It shows images of the simulation with field lines plotted in the vicinity of the fan footpoints (determined in the same manner as {{Figs.~\ref{fig:fan1a}\,--\,\ref{fig:fan1d}}}), side-by-side with SWAP EUV images on the same date, once every two days for each Carrington rotation that the fan is observed, as the fan transits across the Earth-facing solar disc. Coronal magnetic field lines are coloured in the same way as in {{Figs.~\ref{fig:fan1a}\,--\,\ref{fig:fan1d}}}, with the addition of yellow field lines representing occasional disconnected U-loops due to magnetic flux rope ejections.

On 11 October (top row, Fig.~\ref{fig:fan1a}), two fan structures can be seen off the north-east limb in the EUV image. They have the appearance of a bright arc, curving northwards, with a dimmer ``void'' beneath. The fan that we will discuss in detail is the lower of the two, which appears to originate near the Equator and arc up towards the North Pole. Comparing the plotted field lines with the EUV image, the structure of the upper fan appears to have been captured well by the simulation (closed field in particular), but not that of the lower fan. The lower fan will be heavily influenced by active region emergence due to its proximity to the Equator, whereas the upper fan's structure is more likely a consequence of older magnetic flux due to the longer-term surface transport process of meridional flow \citep{duvall1979}. There is a streamer structure in the simulation, composed of closed (dark blue) loops with opposite polarity open field lines on either side (light blue/green), but it is directed radially outward rather than arching towards the North Pole. One of the reasons for this incorrect structure is the late emergence of an active region within the simulation. The active region in question can be seen in the SWAP image, as a bright patch on the east limb, just below the Equator. It can also just be seen on the limb in the HMI magnetogram (see animation HMI\_sim\_compare.mp4). However, it is not assimilated into the simulation until after 13 October, once it has fully rotated onto the Earth-facing side of the Sun and its magnetic flux can be measured.

The emergence of the active region on 13 October has an effect on the structure of the fan within the simulation. This can be seen in Fig.~\ref{fig:fan2}, which shows the simulation on 13 October (before emergence) and 15 October (after emergence), with fan field lines overplotted. The top row shows the Earth-facing side of the Sun. A new active region bipole can be seen in the right-hand image (15 October 2014), just below the fan footpoints (red stars). The bottom row shows the same dates and field lines, with the Sun rotated so that the fan structure can be seen off the limb. In all images, the central meridian as seen from Earth is indicated by a yellow dashed line. It can be seen that the emergence of the new bipole has a significant effect on the connectivity of the fan structure. On 13 October, much of the closed field associated with the fan (dark blue) connects to a slightly decayed bipole north-west of the fan footpoints, towards central meridian. On 15th October, much of this field now connects to the newly emerged bipole instead, so that the large-scale structure is more longitudinally aligned. Another, smaller bipole has emerged in the location of the decayed bipole, creating a small region of closed field (magenta), which also has some effect on the large-scale structure. The right-hand images illustrate that the overall effect of these changes causes the equatorial streamer to arc slightly towards the North Pole, which is more in line with what was observed by SWAP on 11 October. This can also be seen in the animation (fan1\_fl.mp4). The reason why the simulation did not reproduce the lower latitude fan is a consequence of a bipole that emerged on the far side and subsequently could not be included in the model until after its limb transit onto the disk.

On 27 October (bottom row, Fig.~\ref{fig:fan1a}), the fan can be seen off the north-west limb in the SWAP image. The simulation magnetic field lines reproduce the morphology of the fan well, with a closed cusp shape arching up towards the North Pole, open field curving upwards around this, and radially directed open field associated with the active region near the Equator. At this stage, the open magnetic field in the simulation in the vicinity of the fan is predominantly positive (green).

On 8 November (top row, Fig.~\ref{fig:fan1b}), the fan can be seen off the north-east limb in the SWAP image. Again, the simulation magnetic field lines off this limb do not appear to match the structure of the fan well. The large bipole at the centre of disc, north of the Equator, has only just emerged within the simulation, and it is responsible for closed magnetic field (dark blue) underlying the fan structure. It has also introduced some negative open field structure in the vicinity of the fan (light blue). As the simulation progresses, however, there is less closed field associated with this bipole in the vicinity of the fan (this can be seen in the animation, fan1\_fl.mp4), until the structure of the fan is more closely matched by the simulation again on 24 November, when both can be seen off the north-west limb (bottom row, Fig.~\ref{fig:fan1b}). This illustrates that the late assimilation of an active region into the simulation can initially result in incorrect large-scale coronal structures. However, the simulated coronal magnetic field is able to adapt within a few days so that its structure more closely resembles what is observed by the time it reaches the west limb. A pseudostreamer structure can now be seen in the simulation, with two closed field regions, and the open field regions to either side sharing the same polarity as each other. The observed fan appears to be associated with the open magnetic field at the southern side of the pseudostreamer.

On 6 December (top row, Fig.~\ref{fig:fan1c}), the magnetic field structure of the simulation off the north-east limb still matches the observed fan structure quite well, as no large active regions have emerged in the vicinity of the fan footpoints as they rotated around the far side of the Sun. A small active region did emerge at the approximate location of the footpoints, and this is assimilated into the simulation between 8 and 10 December (see the animation, fan1\_fl.mp4 $-$ the new active region bipole can be seen on 10 December). It emerges at the location of the fan footpoints, beneath the closed (dark blue) loops, and it has the effect of expanding the loops' extent, as the bipole's alignment is almost parallel to that of the overlying closed magnetic field.

Several additional new bipoles emerge on the Sun and are incorporated into the simulation as the fan rotates across the Earth-facing solar disc (see animations fan1\_fl.mp4 and HMI\_sim\_compare.mp4). These can be seen on the disc on 20 December (bottom row, Fig.~\ref{fig:fan1c}). The effect that these new emergences have on the EUV observations is that the fan becomes more aligned in the latitudinal direction. A similar effect is seen in the simulation, and the magnetic structure on 20 December is now that of a streamer $-$ a single region of closed loops, with opposite polarity open field on either side. While we are not able to determine the polarity of the coronal magnetic field from the EUV observations, the simulation indicates that the large-scale magnetic structure in the vicinity of the fan may have changed from a pseudostreamer to a streamer.

By 1 January 2015 (top row, Fig.~\ref{fig:fan1d}), the fan has shrunk significantly in size. The large number of active regions that have emerged in the vicinity of the fan's footpoints may have contributed to its reduction in physical extent. The fan appears as a smaller arc structure off the north-east limb in EUV, above the footpoints that have been plotted as red stars. The fan structure within the simulation has a similar, reduced size. On 15 January (bottom row, Fig.~\ref{fig:fan1d}), there are significantly fewer bright, extended features off the north-west limb than there were in the previous rotations (27 October, bottom row, Fig.~\ref{fig:fan1a}; 24 November, bottom row, Fig.~\ref{fig:fan1b}; and 20 December, bottom row, Fig.~\ref{fig:fan1c}). In turn, the simulation shows predominantly closed structures.

\subsection{Coronal Holes}

  \begin{figure}
  
     \centerline{\hspace*{0.0\textwidth}
               \includegraphics[width=0.5\textwidth,clip=]{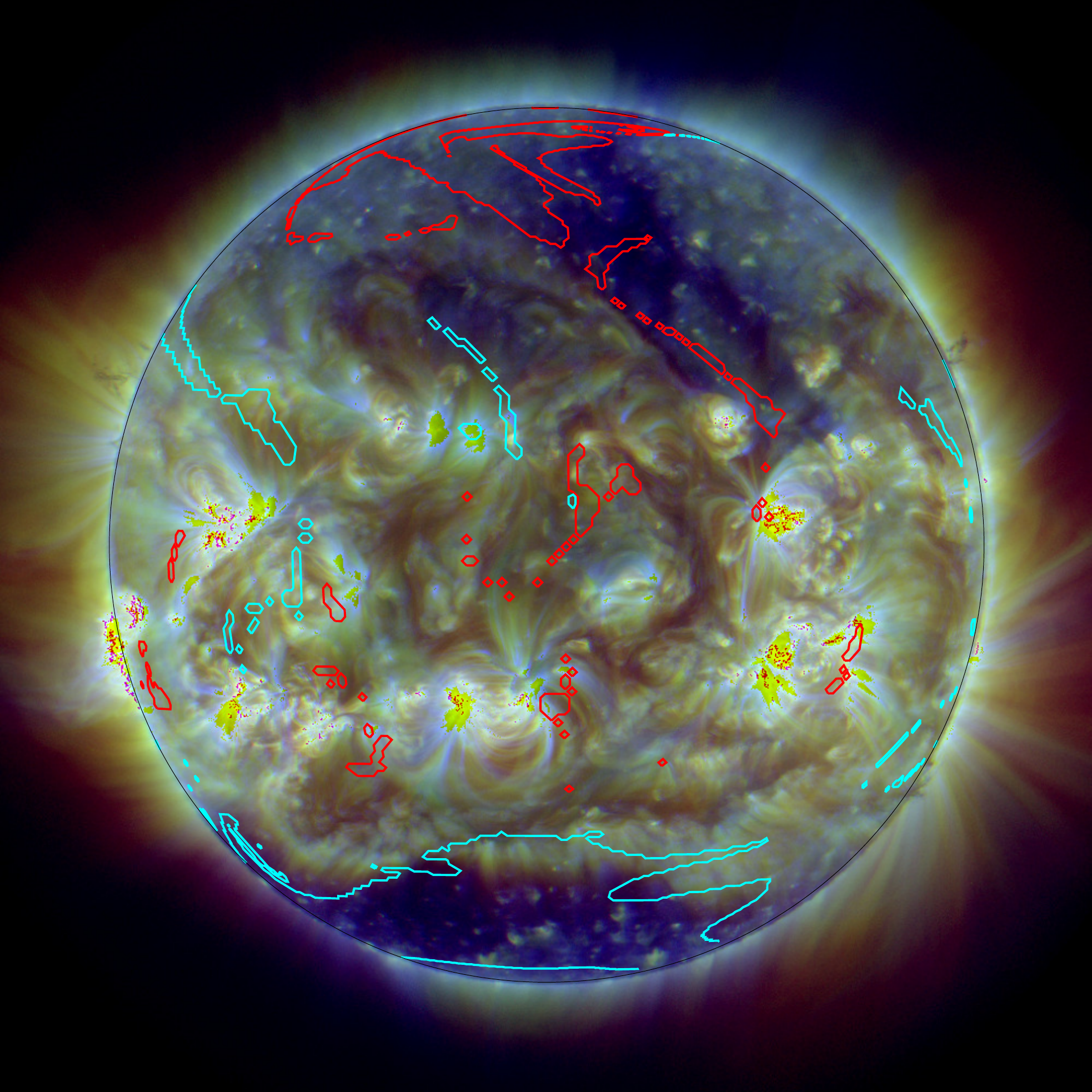}
               \hspace*{-0.0\textwidth}
               \includegraphics[width=0.5\textwidth,clip=]{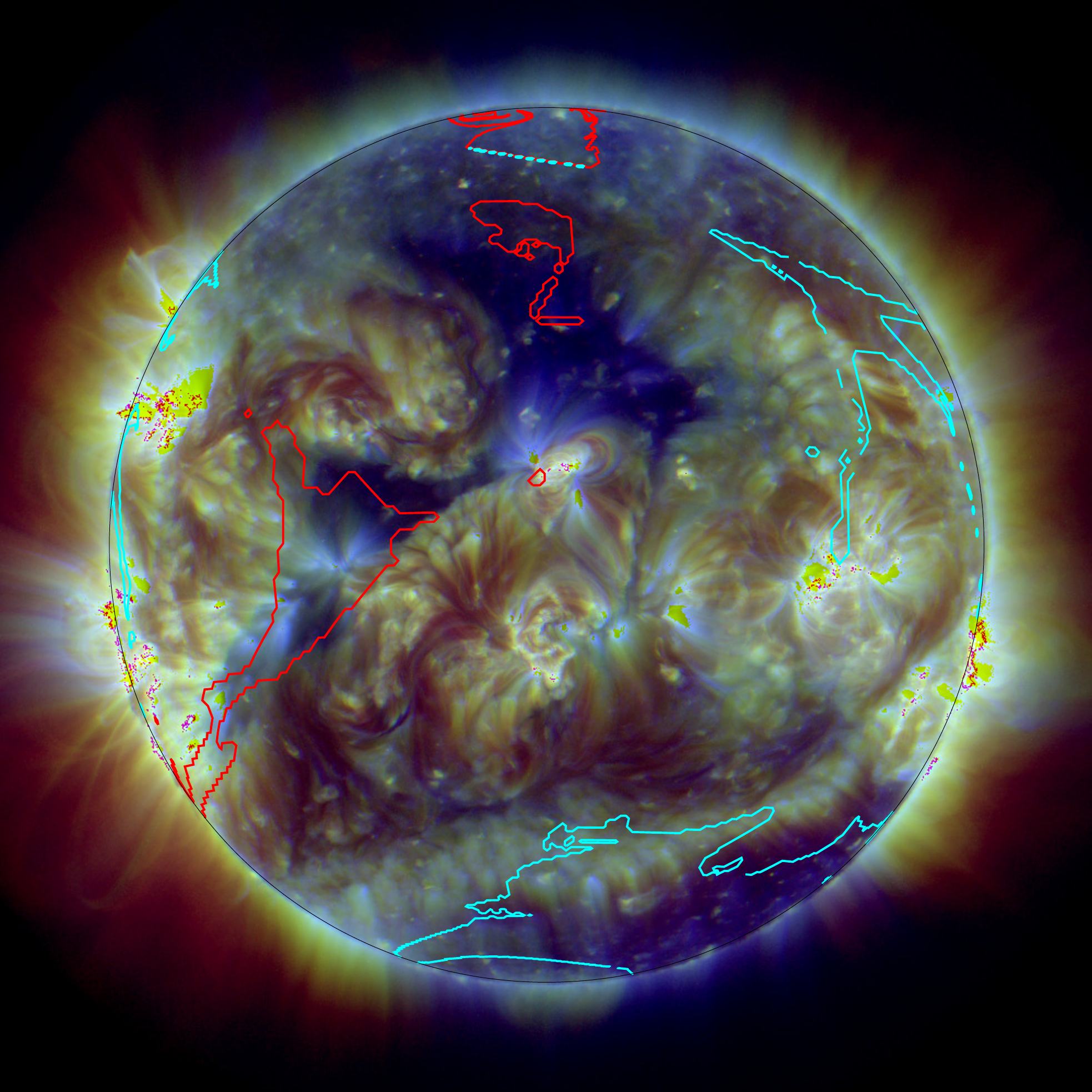}
              }
     \vspace{-0.51\textwidth}   % Shift close to the panel top 
     \centerline{\Large \bf     % Includes the labels (here needs the color 
                                %   package, see beginning of this file)
      \hspace{-0.02 \textwidth}   \color{white}\footnotesize{28-Nov-2014}
      \hspace{0.35\textwidth}  \color{white}{07-Jan-2015}
         \hfill}
     \vspace{0.47\textwidth}    % Shift back to the panel bottom 

\caption{AIA composite images (constructed from 171\,\AA, 193\,\AA\,and 211\,\AA\,observations) overlaid with contours of open magnetic field determined from the simulation on the same date. The contour regions indicate the photospheric footpoints of open magnetic field lines, with red indicating positive and {{light blue}} indicating negative open field.}\label{fig:ch}
   \end{figure}
   
Regions of open magnetic field in the simulation are determined by tracing magnetic field lines from the photosphere and determining whether the endpoint of the field line is at the source surface at 2.5\,$\rsun$. The polarity of the field line's footpoint in the photosphere is also checked, so that regions may be identified as being positive or negative open field. Figure~\ref{fig:ch} shows example AIA composite EUV images with contours of open magnetic field from the simulation overlaid (red$=$positive, {{light blue}}$=$negative open field), from 28 November 2014 and 7 January 2015. The contoured regions indicate the photospheric footpoints of open magnetic field lines. The AIA composite images are constructed from 171\,\AA, 193\,\AA\,and 211\,\AA\,observations. Coronal hole regions in the composite images appear particularly dark compared to the surrounding coronal plasma, allowing for easier visual comparison with the simulation open field contours.

In some cases, the simulation open field regions appear to match the structure of observed coronal hole regions quite well. For example, a large coronal hole can be seen at the South Pole in the AIA images throughout most of the period, and the simulation contours in this region are generally of a similar shape and size, largely overlapping the observed coronal hole. The triangular shaped region of positive (red) open field, lying across the Equator on 7 January, also reproduces the underlying observed coronal hole region reasonably well, although this is not a full coronal hole, due to the presence of fan-like loop structures.
In other cases, open field regions appear offset slightly from the true location or are not of the correct scale. Some examples of this can be seen on 28 November 2014 and 7 January 2015 in the northern half of the disc. In both cases, regions of positive (red) open field in the simulation are too small compared to the observed underlying coronal holes. One reason for the discrepancies is the late assimilation of emerging far-side active regions into the simulation. \cite{weinzierl2016} showed that when active region emergence is delayed by only a few days within the simulation, it can result in significant differences in magnetic connectivity, which is an effect that is not just seen local to the active region itself, but across the global corona. Another potential cause of discrepancies is the smooth nature of the simulated photospheric magnetic field compared to observed magnetograms (e.g. compare the simulation photosphere with the observed HMI magnetograms in Fig. \ref{fig:B}). In a follow-on study we will conduct a detailed comparison between regions of open magnetic field in the simulation and coronal hole regions detected and tracked with the SPoCA-suite \citep{verbeeck2014}. We will also consider the effect on the simulated corona of active region decay through fragmentation, convection and other smaller-scale processes, as opposed to decay through diffusion, which occurs in the current flux transport model.

\section{Discussion and Conclusions}\label{s:conc} 

We have presented a comparison of a global coronal magnetic field simulation with SWAP EUV images from the same dates, focussing in particular on large-scale, extended structures observed off the solar limb. Generally, the simulation reproduces large-scale coronal structures observed off the west limb more accurately than those observed off the east limb, due to our current inability to incorporate far-side emerging active regions into the simulation until after they have rotated onto the Earth-facing side of the Sun. This is demonstrated in particular in the case study of a fan that was observed over several Carrington rotations. The simulation is able to reproduce the polar evolution of the Sun's magnetic field to an extent, since the flux transport model includes the poleward transport of magnetic flux due to meridonal flow. This was seen in the comparison of the cavity/pseudostreamer structure at the South Pole \citep{guennou2016} with the corresponding maximum height of closed magnetic field at the South Pole in the simulation. A decrease in height of the structure over time is seen in both the observations and the simulation, although the simulation does not produce the correct scale of the structure. This is in part due to the limited duration of the simulation, which is seven months in total, compared to the typical timescale for flux transport from the Equator to the poles by meridional flow, which is two years \citep{mackay2012}. Longer term (e.g. years) simulation of the global coronal magnetic field can therefore improve the accuracy of the simulated polar regions \citep{yeates2012}, in particular, the balance of potential/non-potential magnetic field.

There are several possible improvements that could be made to increase the accuracy of the global magnetofrictional model, including additional data products and the development of the model itself. 
Currently, active regions that have emerged on the far-side of the Sun cannot be incorporated into the simulation until they have rotated onto the near-side. Additional magnetogram observations from a spacecraft positioned at $L_5$, such as the proposed ESA \emph{Lagrange} mission, \citep{kraft2017}, could greatly reduce the error associated with this. \cite{mackay2016} have shown that such observations  could increase the accuracy of global quantities in solar coronal simulations by 26\,\%\,--\,40\,\%. \cite{weinzierl2016} showed that a delay in active region emergence of a few days within continuous non-potential simulations (due to limited field-of-view) can result in significant differences in magnetic connectivity, not just local to the active region itself, but also globally. Inaccuracies in predictions of open magnetic field, for example, could significantly impact the prediction of coronal hole regions, solar wind, and the interplanetary magnetic field.

Another product that could be complementary to magnetograph data from $L_5$ is far-side detection of active regions using helioseismology \citep{gonzalez2007}. \cite{liewer2014} showed that over a period of nine months, the \emph{Global Oscillation Network Group} (GONG) predicted 55\,\% of far-side emerging active regions before they rotated to the Earth-side, while HMI predicted 48\,\%. An $L_5$ magnetograph would significantly reduce the number of missed regions and could provide much more detailed information on them, such as the spatial distribution of magnetic flux and orientation of polarities. Solar Orbiter will also produce occasional far-side magnetograms, as well as invaluable observations of polar magnetic fields during its higher latitude orbits, providing further insight for model development \citep{solanki2019}. Such additional information could greatly enhance the accuracy of global solar simulations.

A further possibility for the enhancement of far-side active region detections and properties is the use of machine learning. \cite{chen2019} trained a neural network to ``learn'' the relationship between near-side EUV observations from AIA and magnetic flux observations from HMI. The trained network was then applied to far-side EUV images from four years worth of STEREO/EUVI data to produce far-side magnetic flux maps of the Sun. They have since extended this work by training a further neural network to learn the relationship between the far-side flux maps and far-side acoustic maps determined using helioseismology. The goal is then to apply this second trained network to current helioseismic data, in order to produce far-side magnetic flux maps (since there is only a limited period for which the STEREO spacecraft were observing the far-side of the Sun). The properties of the active regions determined from such far-side predicted magnetic flux maps could be further constrained once they have rotated into the field-of-view of an $L_5$ magnetograph, potentially greatly enhancing our space weather prediction capabilities.

Further possible sources of discrepancy between the model and observations include i) the expansion factor of the magnetic field with radial height; ii) the absence of plasma within the model; iii) errors in the photospheric flux distribution (discussed below); iv) errors in the energisation of the corona through the self-helicity of the bipoles; and v) missing energisation of the corona due to the process of small scale helicity injection such as occurs in the helicity condensation model \citep{antiochos2013,mackay2014}. A future study could compare the effect and relative importance of each of these on the accuracy of global coronal modelling.

Considering the photospheric flux distribution, one limitation of the model discussed in this article is that new active region bipoles are only inserted into the model on the day of their peak observed flux, so that only the decay phase of an active region is simulated. {{Currently, when bipoles are added to the 3D simulation, this is carried out as a simplified process where: i) the main simulation is switched off, ii) a 3D dome for the bipole is created by sweeping away pre-existing flux to produce a magnetic vacuum, iii) the 3D bipole field is added as an isolated system, and iv) the field is allowed to relax and connections are made to the surrounding fields (see \cite{mackay2006} and \cite{yeates2008} for further details). Once this has been carried out, the main simulation is allowed to continue. At the present time stages i) – iv) are carried out in a time-independent manner, therefore the bipole can only be added once. Since it can only be added once, it is best to add it at peak flux.
The above process is much simplified, so future developments to the model will include a time-dependent emergence process. However, it is important to first ensure that this process produces a realistic 3D coronal field. The model in its present form can simulate the decay of active regions and the transport of flux poleward over a period of years. Therefore, it can be used for predictions as long as the prediction is made on the timescales of weeks to months and an adaptive bipole input process is continually applied each day.”}}

One of the advantages of global magnetofrictional simulations is that they are computationally inexpensive enough to be carried out in real time, so that they can provide an energised lower boundary condition for space weather modelling and prediction (e.g. WSA-ENLIL, \citealp{wang1990,wang1995,arge2000,pizzo2011}). Therefore, developments to the model itself should aim to retain this property.
A further development that is planned for the flux transport component of the global evolution model will focus in particular on the decay phase of active regions. Currently, the dispersal of active region flux within the flux transport model is approximated by a diffusion coefficient (Eqs.~\ref{eq:eqn1} and \ref{eq:eqn2}). Observations show that small-scale processes appear to play a role in the dispersal and decay of solar active regions, for example through cancellation at the boundary of sunspot outflows \citep{kubo2007,kubo2008} and erosion of flux by surrounding supergranulation \citep{dacie2016}. The effects of this can be seen in Fig.~\ref{fig:B}, where the flux transport model (Fig.~\ref{fig:B}a\,--\,c) is able to reproduce the general large-scale evolution of the observed photospheric magnetic field (Fig.~\ref{fig:B}d\,--\,f). However, it can be seen that the model photospheric magnetic field is smooth, whereas the observed magnetic field is fragmented. In a follow-up study, we will examine the effect on the coronal evolution of an active region that has decayed as the result of such smaller-scale processes, compared to the standard flux transport model decay of an active region through diffusion. Active region decay via smaller scale processes will be simulated by coupling the current flux transport model to the Magnetic Carpet model of \cite{meyer2011} and \cite{meyer2016}, which successfully reproduced many observed properties of the small-scale photospheric magnetic field.

%%%%%%%%%%%%%%%%%%%%%%%%%%%%%%%%%%%%%%%%%%%%%%%%%%%%%%%%%%%%%%%%%%%%%%%%%%%
%% Appendix
%

%%%%%%%%%%%%%%%%%%%%%%%%%%%%%%%%%%%%%%%%%%%%%%%%%%%%%%%%%%%%%%%%%%%%%%%%%%%
%% Acknowledgements
%
\begin{acks}
K.A. Meyer gratefully acknowledges the support of the PROBA2 Guest Investigator Program, the Carnegie Trust for the Universities of Scotland and the STFC. K.A. Meyer would like to thank Marilena Mierla for help with the SWAP data, and Huw Morgan for assistance in enhancing the faint off-limb features observed in EUV (eagle.imaps.aber.ac.uk/About.html). D.H. Mackay would like to thank both the UK STFC and the ERC (Synergy grant: WHOLE SUN, grant Agreement No. 810218) for financial support. D.H. Mackay would like to thank STFC for IAA funding under grant number SMC1-XAS012. D.C. Talpeanu was funded by a Ph.D. fellowship of the Research Foundation –
Flanders (FWO), and was partially supported by a PhD Grant from the Royal Observatory of Belgium and from the Belgian Federal Science Policy Office (BELSPO). L.A. Upton was supported by the National Science Foundation Atmospheric and Geospace Sciences Postdoctoral Research Fellowship Program (Award Number:1624438). SWAP is a project of the Centre Spatial de Li\`ege and the Royal Observatory of Belgium funded by the Belgian Federal Science Policy Office (BELSPO). The authors are grateful to the SDO/HMI and AIA teams for the data, which was obtained through the Joint Science Operations Centre (JSOC: jsoc.stanford.edu). SWAP data preparation routines were used within SolarSoft IDL (www.lmsal.com/solarsoft/). We thank the anonymous reviewer for their comments and suggestions, which have improved the article. 
\end{acks}

\section*{Disclosure of Potential Conflicts of Interest}
The authors declare that they have no conflicts of interest.
%%% %%%%%%%%%%%%%%%%%%%%%%%%%%%%%%%%%%%%%%%%%%%%%%%%%%%%%%%%%%%
%% Bibliography
%
% Using BibTeX
%
 \bibliographystyle{spr-mp-sola}
 \bibliography{swap_paper_submit_v4_no_blue}  
%
% Without BibTeX 
% \begin{thebibliography}{}
% \bibitem[\protect\citeauthoryear{Author}{Year}]{key}
%   <bibliographical entry>
%
% \bibitem[\protect\citeauthoryear{}{}]{}
%   
%  
% \end{thebibliography}

\end{article} 
\end{document}